\documentclass[fleqn,usenatbib]{mnras}

\usepackage[T1]{fontenc}
\usepackage{ae,aecompl}

\usepackage{graphicx}    
\usepackage{amsmath}    
\usepackage{amssymb}    
\usepackage{stfloats}
\usepackage{adjustbox}

\title[Unsupervised Galaxy Morphological Classification]{Beyond the Hubble Sequence - Exploring Galaxy Morphology with Unsupervised Machine Learning}

\author[Ting-Yun Cheng et al.]{Ting-Yun Cheng,$^{1,2}$\thanks{E-mail:ting-yun.cheng@nottingham.ac.uk}
Marc Huertas-Company,$^{3,4}$
Christopher J. Conselice,$^{2,5}$
\newauthor
Alfonso Arag\'on-Salamanca,$^{2}$
Brant E. Robertson,$^{6}$
Nesar Ramachandra$^{7}$
\newauthor
\\ \\
$^{1}$ Centre of Extragalactic Astronomy, Durham University, Stockton Rd, Durham DH1 3LE, UK\\
$^{2}$ School of Physics and Astronomy, University of Nottingham, University Park, Nottingham, NG7 2RD, UK\\
$^{3}$ Instituto de Astrof\'isica de Canarias (IAC); Departamento de Astrof\'isica, Universidad de La Laguna (ULL), E-38200, La Laguna, Spain\\
$^{4}$ LERMA, Observatoire de Paris, CNRS, PSL, Universit\'e Paris Diderot, France\\
$^{5}$ Jodrell Bank Centre for Astrophysics, University of Manchester, Oxford Road, Manchester UK\\
$^{6}$ Department of Astronomy and Astrophysics, University of California, Santa Cruz, 1156 High Street, Santa Cruz, CA 95064 USA\\
$^{7}$ High Energy Physics Division, Argonne National Laboratory, Lemont, IL 60439\\
}

\date{Accepted 2021 March 8. Received 2021 March 5; in original form 2020 September 28}

\pubyear{2021}

\begin{document}
\label{firstpage}
\pagerange{\pageref{firstpage}--\pageref{lastpage}}
\maketitle

\begin{abstract}
We explore unsupervised machine learning for galaxy morphology analyses using a combination of feature extraction with a vector-quantised variational autoencoder (VQ-VAE) and hierarchical clustering (HC). We propose a new methodology that includes: (1) consideration of the clustering performance simultaneously when learning features from images; (2) allowing for various distance thresholds within the HC algorithm; (3) using the galaxy orientation to determine the number of clusters. This setup provides 27 clusters created with this unsupervised learning which we show are well separated based on galaxy shape and structure (e.g., S\'ersic index, concentration, asymmetry, Gini coefficient). These resulting clusters also correlate well with physical properties such as the colour-magnitude diagram, and span the range of scaling-relations such as mass vs. size amongst the different machine-defined clusters. When we merge these multiple clusters into two large preliminary clusters to provide a binary classification, an accuracy of $\sim87\%$ is reached using an imbalanced dataset, matching real galaxy distributions, which includes 22.7\% early-type galaxies and 77.3\% late-type galaxies. Comparing the given clusters with classic Hubble types (ellipticals, lenticulars, early spirals, late spirals, and irregulars), we show that there is an intrinsic vagueness in visual classification systems, in particular galaxies with transitional features such as lenticulars and early spirals. Based on this, the main result in this work is not how well our unsupervised method matches visual classifications and physical properties, but that the method provides an independent classification that may be more physically meaningful than any visually based ones.

\end{abstract}


\section{Introduction}

Galaxy structure and visual morphology have a strong connection with their stellar population properties, such as surface brightness, colour, and the formation history of galaxies \citep{Holmberg1958, Dressler1980}. The dominant visual morphological classification system in use today was first constructed by \citet{Hubble1926}, which was then revised by adding a class for lenticulars (S0), a type of galaxy has a disk structure without apparent spiral arms \citep{Hubble1936, Sandage1961}. Since then, a number of detailed classification systems were proposed such as ones including the notation for the inter and outer ring structure \citep{deVaucouleurs1959} and different arm classes \citep{Elmegreen1982, Elmegreen1987}, among others.

However, visual classification systems can be intrinsically biased due to the subjective judgement of different human classifiers. These human errors are unavoidable and sometimes cannot be reproduced for carrying out a statistical analysis. This greatly limits the ability to use galaxy classification in a formal quantitative way. These issues led astronomers to search for a quantitative description of galaxy structure based on the shape, structure, and physical properties of galaxies which can in principle be connected with visual morphology. For example, Principal Component Analysis (PCA) was applied to determine the number of dominant features needed to reproduce the variance shown in observation in \citet{Whitmore1984} as well as to provide an objective procedure for analysing galaxy properties \citep[also see][]{Conselice2006}. Other studies such as non-parametric methods, e.g., concentration, asymmetry, smoothness/clumpiness, and Gini coefficient \citep[][]{Conselice2000, Bershady2000, Abraham2003, Conselice2003, Lotz2004, Law2007}, and parametric methods, e.g., S\'ersic profile \citep{sersic1963, sersic1968} for measuring galaxy structure were also proposed to provide a more objective and quantitative classification systems than visual assessment alone.

Even though quantitative measures of galaxy structure are extremely useful for measuring properties such as the merger history \citep[e.g.,][]{Conselice2003}, morphological `classifications' into types is still an important and complementary process. However, it is not clear if indeed we know what these `best types' are such that whether a classification scheme results in relatively unique physical properties of the galaxies or traces the merger history in each class.

Thus, in this study we build a galaxy morphological classification system that does not involve human bias; we do this through an unsupervised machine learning approach. One may argue that supervised machine learning, as a more established technique, might be more suitable for this task. However, providing labelled data, which is essential to train a supervised machine, must involve human judgement. Our intention in this study is to avoid human bias. Second, the high accuracy achieved by a supervised machine is based on the given prior knowledge from labelled data. Therefore, the performance of a supervised machine becomes uncertain when the testing domain is different from the training domain \citep[e.g.,][]{Dodge2016, Rosenfeld2018}. In future surveys such as the Vera Rubin Observatory \citep[formerly known as Large Synoptic Survey Telescope, LSST;][]{Ivezic2019}\footnote{https://www.lsst.org} will generate the size of the Sloan Digital Sky Survey \citep[SDSS;][]{York2000}\footnote{https://www.sdss.org} data over ten years in one night. It is doubtful that the current data and human labelling abilities and speed could facilitate an unbiased and extensive labelled dataset for a supervised machine to `correctly' classify the potentially unknown patterns within the data of future surveys.  Furthermore, this supervised learning will always miss unusual systems.

For the reason discussed above, we use unsupervised machine learning which is trained without any prior knowledge (e.g., galaxy labels, such as Hubble types). This approach is able to give us suggestive classifications from the machine's perspective based upon input features. However, with an unsupervised machine learning technique it becomes more challenging to have a `sensible' classification, that is one with more consistency with human opinion, when the dimensionality of a feature space becomes high \citep[curse of dimensionality,][]{Bellman1954, Keogh2017}. In astronomical studies, unsupervised machine learning applications have been mostly used in the studies of spectroscopic data which is less dimensional than applying to imaging data \citep[e.g.,][]{Geach2012, Krone-Martins2014, Carrasco2014, Siudek2018}.  Therefore, unsupervised learning for galaxy classification is still in its infancy.

There are currently several types of astronomical studies that apply unsupervised machine learning techniques to images which reach reasonable results, including: galaxy morphology \citep{Hocking2018, Martin2020}, strong lensing identification \citep{Cheng2020b}, and anomaly detection \citep{xiong2018, Margalef-Bentabol2020}. For example, \citet{Hocking2018} and \citet{Martin2020} apply a technique called Growing Neural Gas algorithm \citep{Fritzke1994}, which is a type of Self-organising Map \citep[SOM,][]{Kohonen1997}, to extract features from images.  These features are then connected with a hierarchical clustering algorithm \citep{Hastie2009}. On the other hand, \citet{Cheng2020b} use a fundamentally different approach by using a convolutional autoencoder \citep{Masci2011}, which includes an architecture of convolutional neural networks, for feature extraction. This method connects the extracted features with a Bayesian Gaussian mixture model from which a clustering analysis can be done.

In this study, we apply an architecture consisting of a convolutional autoencoder, as convolutional neural networks have demonstrated their capability for capturing representative and meaningful features from images \citep{Krizhevsky2012}. We do not use the same convolutional autoencoder as \citet{Cheng2020b}, but we apply a newly developed technique from Google DeepMind \citep{vandenOord2017, Razavi2019} called `Vector-Quantised Variational Autoencoder (VQ-VAE)'. This technique includes a vector quantisation method that accelerates the time-consuming process of feature extraction when using a convolutional autoencoder, as explained in \citet{Cheng2020b}. On the other hand, for clustering algorithms, we decide to apply a modified hierarchical clustering method to group the data in order to explore connections between the distances amongst extracted features in feature space, and the number of classification clusters.

In this paper, we use this unsupervised machine learning technique to develop a galaxy morphology classification system defined by a machine, and compare it with traditional visual classification system such as the Hubble sequence.  We furthermore also compare our machine developed classification with galaxy physical properties, such as stellar mass, colour, and physical size of galaxies.  We use monochromatic images throughout to focus only on the impact of galaxy shape and structure on morphological classifications in this paper. The methodology we develop is introduced in Section~\ref{ch6_sec:methods}, while the detailed description of how to approach using our method and the data used in this study are shown in Section~\ref{ch6_sec:implementation}. Section~\ref{ch6_Sec:result} presents the results in this study. Finally, we conclude the work in Section~\ref{ch6_sec:conclusion}.

\section{Methodology}
\label{ch6_sec:methods}

In this section we explain our unsupervised machine learning methodology that is used throughout this paper.  We give a brief overview here, before going into detail in the following subsections.

Our unsupervised machine learning technique includes a feature learning phase with a vector-quantised variational autoencoder (VQ-VAE; Section~\ref{ch6_sec:vqvae} and Section~\ref{ch6_sec:modi_vqvae}) and a clustering phase using a hierarchical clustering algorithm (HC; Section~\ref{ch6_sec:hc}). Several novel approaches for unsupervised machine learning applications are made in this paper: (1) the VQ-VAE considers both reconstruction and preliminary clustering results in the feature learning phase (Section~\ref{ch6_sec:modi_vqvae} and also see Section~\ref{ch6_Sec:feature_learning}); (2) multiple different distance thresholds are used to draw the decision lines on the merger tree in the clustering process (Section~\ref{ch6_sec:hc}); (3) use the feature of galaxy orientation to decide the distance thresholds applied in the clustering process (see details in Section~\ref{ch6_sec:hc}).

\subsection{Vector-Quantised Variational Autoencoder (VQ-VAE)}
\label{ch6_sec:vqvae}

\begin{figure*}
\begin{center}
\graphicspath{{figures/}}
	\includegraphics[width=2\columnwidth]{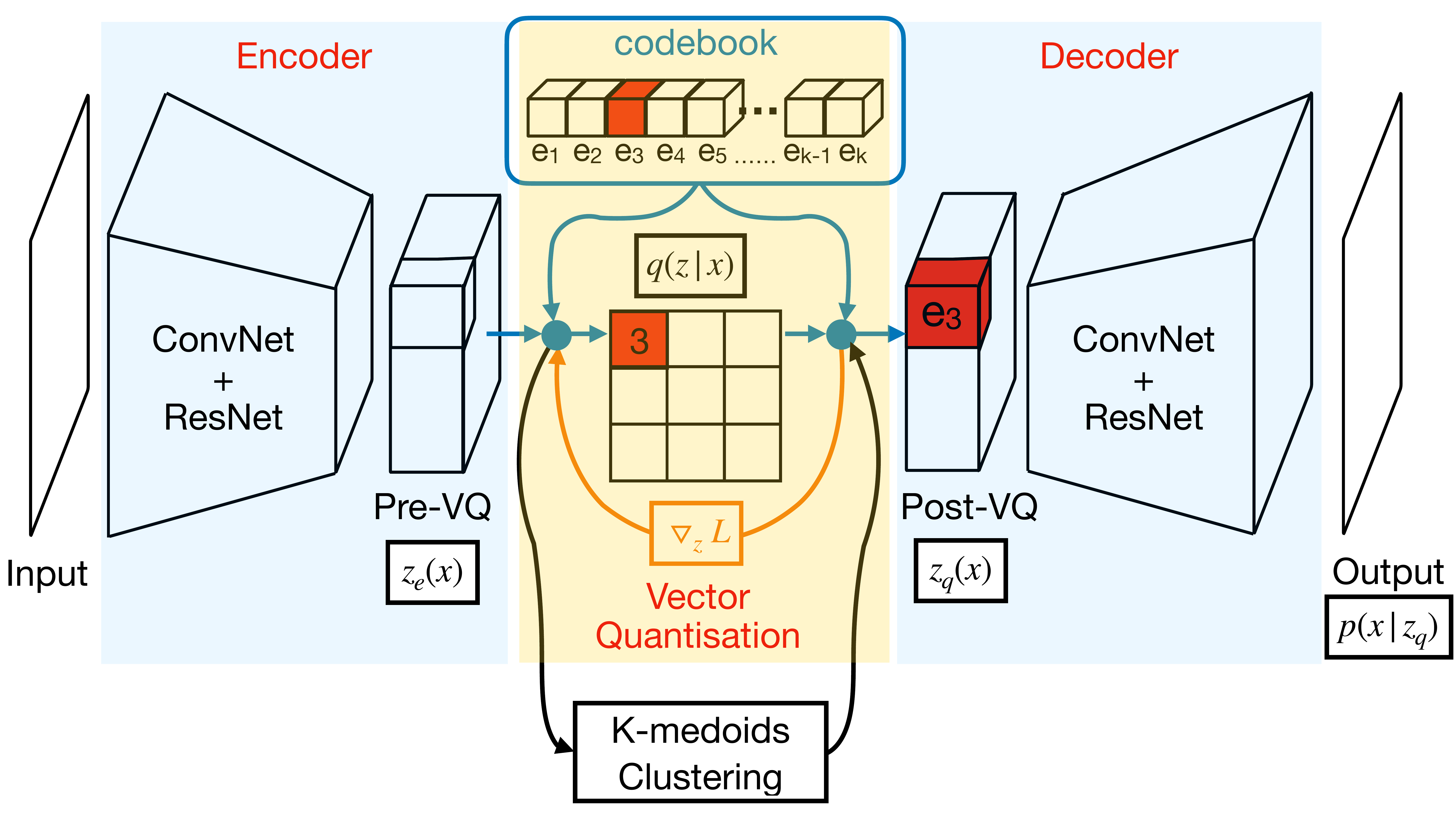}
   	\caption{A schematic architecture of the {\bf modified} VQ-VAE used for feature extraction of images. The top aspect with a coloured background is the main architecture of the VQ-VAE, which is then modified to consider the silhouette score calculated using the two preliminary clusters given by k-medoids clustering as a part of the loss function when training VQ-VAE (see details in Section~\ref{ch6_sec:modi_vqvae}). The blue shading at the left and right represents the encoder and the decoder, respectively while the yellow part shows the vector quantisation process. The details of each layer are shown in Table~\ref{ch6_tab:vqvae}}
    \label{ch6_fig:vqvae}
\end{center}
\end{figure*}

The vector-quantised variational autoencoder (VQ-VAE) was built by Google DeepMind \citep{vandenOord2017, Razavi2019} and was originally used for high-fidelity image emulation. The task of image emulation is to learn the distribution of the data given a set of training images, and then to reproduce the images with the learnt distribution. In details, the structure of an autoencoder (Fig.~\ref{ch6_fig:vqvae}) contains an encoder with a posterior distribution $q\left( { z }|{ x } \right) $ and a prior distribution $p\left( { z } \right) $ where $x$ is the input data and $z$ represents latent variable, and a decoder with a distribution $p\left( { x }|{ z } \right) $ for reproducing the input data.

The VQ-VAE is a type of autoencoder which includes the structure of convolutional neural networks and applies a vector quantisation process \citep{vandenOord2017} to make the posterior and prior distribution become categorical. By using a categorical distribution, the computational time for training an autoencoder is significantly reduced compared to other machine learning methods. For example, in \citet{Cheng2020b}, it takes 0.0146 milliseconds (ms) per pixel per epoch by a convolutional autoencoder running on a NVIDIA TU102 [GeForce RTX 2080 Ti] GPU, while in this work, a VQ-VAE takes $4.59\times10^{-4}$ ms per pixel per epoch using the same device. This is an enormous difference in training speed (30 times faster), but without reducing the reconstruction ability.  The reconstruction errors of both methods are on average $\sim5\times10^{-4}$ when using the most optimal trained model. This shows the usefulness of a built-in vector quantisation process to an autoencoder. More importantly, the improvement in computational efficiency makes it feasible to apply such unsupervised techniques to large-scale survey data, even allowing it to be included in the analysis pipelines for future surveys.

Following the top coloured area in Fig.~\ref{ch6_fig:vqvae}, the posterior categorical distribution $q\left( { z }|{ x } \right) $ is defined as \citep{vandenOord2017, Razavi2019}:
\begin{equation}
\label{eq:vq_1}
    	q\left( z={ k }|{ x } \right) =\begin{cases} 1\quad for\quad k={ {\rm argmin} }_{ j }{ \left\| { z }_{ e }\left( x \right) -{ e }_{ j } \right\|  }_{ 2 } \\ 0\quad otherwise \end{cases},
\end{equation}
where ${ z }_{ e }\left( x \right)$ is the output of the encoder (the blue part at the left in the figure), the value ${ e }_{ j }$ represents a vector in the codebook which is used for vector-quantising the ${ z }_{ e }\left( x \right)$, and $k$ is the index for the vector used in the selected codebook (the top box of the yellow part in the figure). We then measure the vector-quantised representation ${ z }_{ q }\left( x \right)$, which is the input of the decoder (the blue shading at the right side in the figure), through Equations~\ref{eq:vq_1} and \ref{eq:vq_2}.
\begin{equation}
\label{eq:vq_2}
    	{ z }_{ q }\left( { x } \right) ={ e }_{ k },\quad {\rm where}\quad k={ {\rm argmin} }_{ j }{ \left\| { z }_{ e }\left( x \right) -{ e }_{ j } \right\|  }_{ 2 }.
\end{equation}
\noindent The vector quantisation process is shown as the yellow part in Fig.~\ref{ch6_fig:vqvae}. The output of an encoder, ${ z }_{ e }\left( x \right)$ can be represented by a combination of the index of different vectors, $k$, in the codebook (the square in the middle of the yellow part). For example, in Fig.~\ref{ch6_fig:vqvae}, a voxel in the output of an encoder is represented by a vector, $e_{3}$, after the vector quantisation. We then use the index of these vectors to build a two dimensional index map.  For the pixel used in our example the value is 3.   With this index map, we can rebuild the distribution, ${ z }_{ q }\left( x \right)$, with the same dimension as ${ z }_{ e }\left( x \right)$ but in this case each `pixel' in ${ z }_{ q }\left( x \right)$ is quantised to one of the vectors shown in the codebook.  For our example, the vector $e_{3}$ is used for the pixel. The distribution of ${ z }_{ q }\left( x \right)$ is then used as the input for the decoder to reconstruct the images.

The loss function of the original VQ-VAE contains three parts: reconstructed loss, codebook loss, and commitment loss. An additional penalty is considered later in the modified version of the VQ-VAE (see Section~\ref{ch6_sec:modi_vqvae}). The reconstructed loss is measured by comparing the reconstructed images with the input images. The codebook loss is used to make the selected codebook, $e_j$, approach the output of the encoder, ${ z }_{ e }\left( x \right)$, while the commitment loss is applied to encourage the ${ z }_{ e }\left( x \right)$ to be as close as possible to the chosen codebook from the previous epoch. With these definitions, the loss function, $L$, for the VQ-VAE is defined as \citep{Razavi2019}:
\begin{equation}
\label{eq:vq_loss}
    	L=\log { p\left( { x }|{ { z }_{ q }\left( { x } \right)  } \right) +{ \left\| sg\left[ { z }_{ e }\left( { x } \right)  \right] -e \right\|  }_{ 2 }^{ 2 }+\beta { { \left\| { z }_{ e }\left( { x } \right) -sg\left[ e \right]  \right\|  }_{ 2 }^{ 2 } } },
\end{equation}
\noindent where the value $sg$ is the stopgradient operator and $\beta$ is used for adjusting the weight of the commitment loss. The study of \citet{vandenOord2017} found that their results correlate with the value of $\beta$, and no apparent change occurs when $\beta$ ranges from 0.1 to 2.0.  Therefore, we set $\beta=0.25$ in this study which follows the setting in \citet{vandenOord2017}.

The details of the VQ-VAE architecture is shown in Table~\ref{ch6_tab:vqvae}. Four convolutional layers are used in both the encoder and decoder, and residual neural networks \citep[ResNets,][]{He2016} are used in this architecture to create a deeper neural network with less complexity. The activation function applied in the convolutional layers is the Rectified Linear Unit ({\texttt{ReLu}}) \citep{Nair2010_cs} such that $f(z)=0$ if $z<0$ while $f(z)=z$ if $z\ge0$. The VQ-VAE code is based upon the example provided in {\sc sonnet} library \citep{sonnet}\footnote{https://github.com/deepmind/sonnet} which is built on top of {\sc TensorFlow} \citep{tensorflow2015-whitepaper}\footnote{https://www.tensorflow.org}. To train the VQ-VAE, we apply the {\texttt{Adam}} Optimiser \citep{Kingma2014} and the learning rate is set to 0.0003 which is used in \citet{Razavi2019}.
\begin{table*}
	\centering
	\begin{tabular}{ccccc} 
		\hline
		\multicolumn{1}{c}{Type} & {\#channel} & {kernel size} & {stride size} & {activation function}\\
		\hline\hline
		\multicolumn{5}{c}{Encoder}\\
		\hline
		\multicolumn{1}{c}{Convolutional layer} & {64} & {4$\times $4} & {2$\times $2} & {ReLu}\\
		\multicolumn{1}{c}{Convolutional layer} & {128} & {4$\times $4} & {2$\times $2} & {ReLu}\\
		\multicolumn{1}{c}{Convolutional layer} & {128} & {4$\times $4} & {2$\times $2} & {ReLu}\\
		\multicolumn{1}{c}{Convolutional layer} & {128} & {3$\times $3} & {1$\times $1} & {ReLu}\\
		\multicolumn{1}{c}{ResNets} & {} & {} & {} & {}\\
		\hline
		\multicolumn{5}{c}{Pre-VQ}\\
		\hline
		\multicolumn{1}{c}{Convolutional layer} & {64} & {1$\times $1} & {1$\times $1} & {}\\
		\hline
		\multicolumn{5}{c}{Decoder}\\
		\hline
		\multicolumn{1}{c}{Convolutional layer} & {128} & {3$\times $3} & {1$\times $1} & {ReLu}\\
		\multicolumn{1}{c}{ResNets} & {} & {} & {} & {}\\
		\multicolumn{1}{c}{Transposed Convolutional layer} & {128} & {4$\times $4} & {2$\times $2} & {ReLu}\\
		\multicolumn{1}{c}{Transposed Convolutional layer} & {64} & {4$\times $4} & {2$\times $2} & {ReLu}\\
		\multicolumn{1}{c}{Transposed Convolutional layer} & {1} & {4$\times $4} & {2$\times $2} & {}\\
		\hline\hline
		\multicolumn{5}{c}{ResNets}\\
		\hline
		\multicolumn{1}{c}{Convolutional layer} & {32} & {3$\times $3} & {1$\times $1} & {ReLu}\\
		\multicolumn{1}{c}{Convolutional layer} & {128} & {1$\times $1} & {1$\times $1} & {ReLu}\\
		\hline
	\end{tabular}
	\caption{The architecture used for the setup of the VQ-VAE used throughout this study.}
	\label{ch6_tab:vqvae}
\end{table*}

\subsection{Modified VQ-VAE}
\label{ch6_sec:modi_vqvae}

In this study, we apply a modification to our original VQ-VAE to consider both image reconstruction and a preliminary clustering result when extracting the representative features from images (Fig.~\ref{ch6_fig:vqvae}). To achieve this goal, a penalty defined by silhouette score \citep[][Equation~\ref{eq:silhouette_score}]{Rousseeuw1987} is added in the original loss of VQ-VAE (Equation~\ref{eq:penalty4cluster}). The silhouette score indicates how well clusters are separated from each other and is defined by the formula,
\begin{equation}
\label{eq:silhouette_score}
    s=\frac{b-a}{{\rm max}\left ( b,a \right )},
\end{equation}
where $a$ represents the mean intra-cluster distance while $b$ is the distance between a cluster and its nearest neighbour cluster. Therefore, a larger silhouette score indicates a better separation between clusters in feature space. To train our VQ-VAE, we minimise the final loss function combining the loss described in Equation~\ref{eq:vq_loss} and the penalty defined as,
\begin{equation}
\label{eq:penalty4cluster}
    L_{s} = \left ( 1-s \right )\lambda,
\end{equation}
where $s$ represents the silhouette score and $\lambda$ is a constant used for making the magnitude of this penalty of the same order as other losses used in the VQ-VAE (Section~\ref{ch6_sec:vqvae}). The value of $\lambda$ is equal to 0.1 in our case.

As shown in Fig.~\ref{ch6_fig:vqvae}, during the training of the VQ-VAE, we interpolate an unsupervised instance-based clustering algorithm called `k-medoid clustering' \citep{Maranzana1963, Park2009} to obtain two preliminary clusters using a flattened index map. The two clusters are then used for measuring a silhouette score to evaluate the performance of the initial clustering. This step is simply to make VQ-VAE intentionally extract features that can not only be used to reconstruct the input images but also be well separated into at least two distinctive groups in feature space. The Hamming distance \citep{Hamming1950} is used as the distance metric as our data is represented by the indices of the vectors in the codebook whereby the number itself only represents a category rather than a real value of the vector (more description in Section~\ref{ch6_sec:hc}). The `k-medoid clustering' is used here for a fast evaluation; in the main clustering process after feature extraction,  we apply hierarchical clustering algorithms (Section~\ref{ch6_sec:hc}).

\subsection{Uneven Iterative Hierarchical Clustering}
\label{ch6_sec:hc}
\begin{figure}
\begin{center}
\graphicspath{{figures/}}
	\includegraphics[width=\columnwidth]{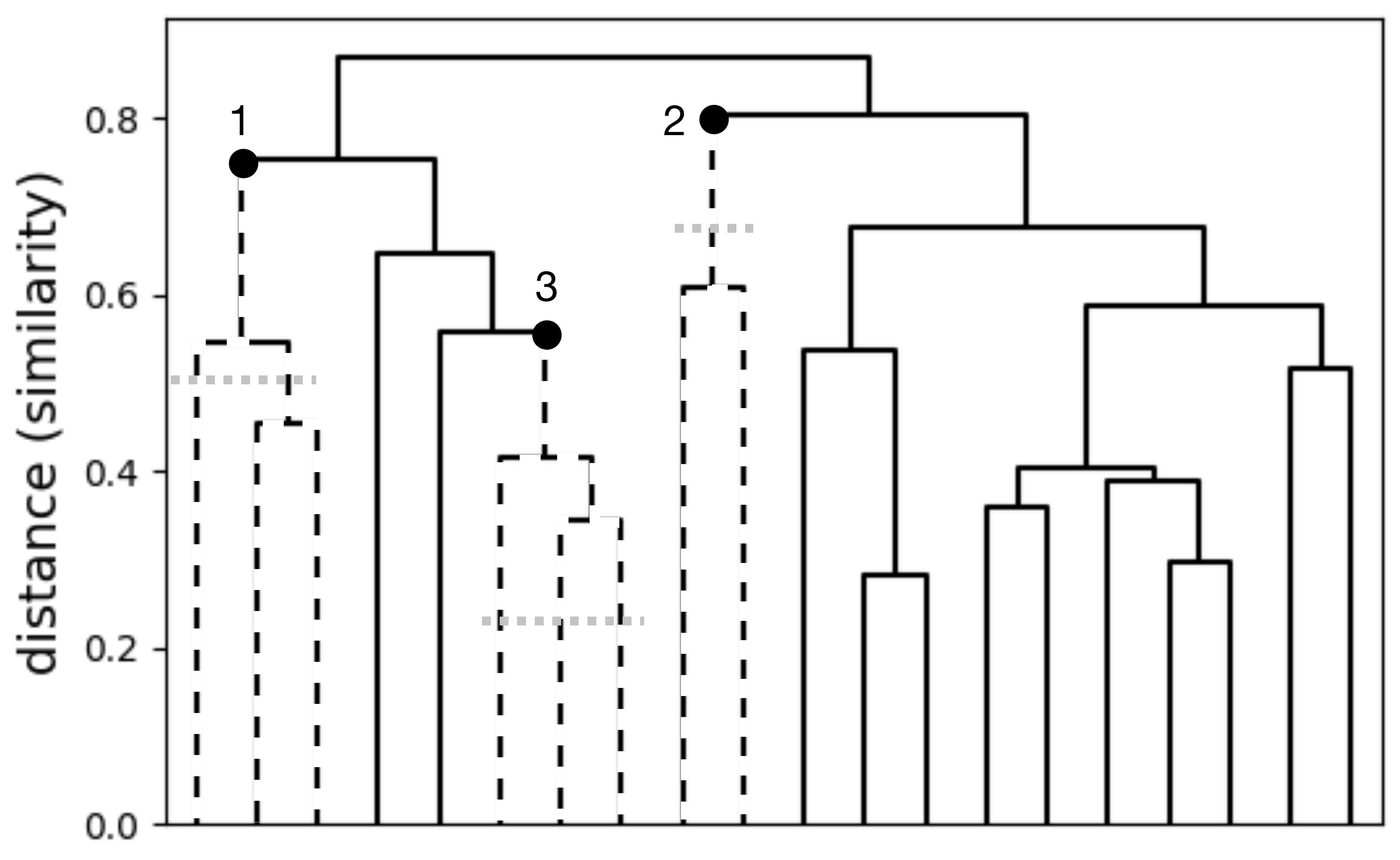}
   	\caption{The schematic dendrogram of the HC process. Datapoints are shown on the $x$-axis, and gradually merge with each other based on the distance (similarity) at the $y$-axis. Each solid line represents a branch and each black circle indicates a stopping point for the corresponding branch (see Section~\ref{ch6_Sec:clustering}). The dashed lines represents the leaves (clusters) after the stopping points. The gray dotted line indicates a cut suggesting the number of clusters in a branch without showing orientation effect (also see Section~\ref{ch6_Sec:clustering}; the results are shown in Section~\ref{ch6_Sec:machine_clf}).}
    \label{ch6_fig:hc}
\end{center}
\end{figure}

In this section we describe our hierarchical clustering procedure for identifying different types of clusters.  Hierarchical Clustering \citep[HC;][]{Johnson1967, Hastie2009}, in particular agglomerative HC (called sometimes `bottom-up'), first assigns each input as an individual group, then merges two nearest (the most similar) groups together based upon the measured pair distance in the feature space, recursively. The `bottom-up' HC structure allows a different number of datapoints in clusters because it starts with individuals (Fig.~\ref{ch6_fig:hc}).  Other kinds of clustering such as `top-down' HC and K-medoid clustering used in Section~\ref{ch6_sec:modi_vqvae} start with clusters themselves, which are more difficult to provide a starting point for an uneven number of datapoints for the initial clusters.

The distance (similarity) measured in this study is the Hamming distance \citep{Hamming1950}. As stated in Section~\ref{ch6_sec:modi_vqvae}, our data is represented by the index of the vectors selected from the codebook.  This is such that an index indicates a category rather than the real value of a vector. We compare two data sets represented by a set of features labelled with indices. The Hamming distance is defined as the number of mismatched indices between the pair over the number of features used to represent the data. For example, assuming that an image can be presented by four different features labelled with the indices: 1, 2, 3, 4, after VQ-VAE; in this case the Hamming distance is 0 if the other image is represented as 1, 2, 3, 4 as well, and the Hamming distance is 1 if it is represented by 4, 3, 2, 1.

For further clarification, Fig.~\ref{ch6_fig:hc} illustrates the clustering process. Within this study, we realise that when all the data are considered, the merging point can be less accurate due to the mixture of blindly measured distances from a great variety of extracted features in images. Therefore, we carry out an iterative clustering process with a reverse concept that we control the data used for doing HC from the top to bottom. We first make the HC merge all data into two top parent branches, then apply the second round of HC to the data of a parent branch to obtain two children branches, and apply the same procedure again to the sub-data of a child branch to get two grandchildren branches, and so on. The iterative action stops when it reaches a certain condition (the black circle in Fig.~\ref{ch6_fig:hc}; see Section~\ref{ch6_Sec:clustering}).

In a typical HC, a uniform distance is used to determine the final clusters. However, a uniform distance threshold is not appropriate considering that galaxies' appearance in different morphological types have different complexity, such that spiral galaxies have a larger diversity in appearance than elliptical galaxies. Therefore, in this study, we propose to allow a different stopping point/distance threshold for each branch depending on the complexity of the objects in the branch (see Section~\ref{ch6_Sec:clustering}). For example, a branch which consists of galaxies which can look very different within a class may continue for many iterations, while others may reach the stop criteria with fewer iterations due to a relatively monotonous structure within the data of the branch.   For example, spiral galaxies can have a variety of spiral arms appearances, i.e., different number of arms, different positions of arms, etc. Therefore, the distance between spiral-like galaxies are generally larger than the distance between two elliptical-like galaxies.  This consideration is sensible and is of great importance in morphological classification of galaxies; however, this is neglected in a typical HC algorithm. Therefore, to distinguish it from a typical HC algorithm, we call this setup `uneven clustering' which provides us with a more precise distinction in galaxy shape, structure, and morphology.

\section{Implementation}
\label{ch6_sec:implementation}

The pipeline of this study includes three main steps: (1) feature selection; (2) feature learning (using the modified VQ-VAE); and finally (3) clustering process. The data used in this study are introduced in Section~\ref{ch6_sec:datasets}. The feature selection is described in Section~\ref{ch6_Sec:feature_selection}, and the setup for the feature learning process using the modified VQ-VAE (Section~\ref{ch6_sec:modi_vqvae}) is discussed in Section~\ref{ch6_Sec:feature_learning}. Finally, in Section~\ref{ch6_Sec:clustering} we explain the details of the clustering process we use to classify galaxies.

\subsection{Data Sets}
\label{ch6_sec:datasets}

The imaging data used throughout this work is from the Sloan Digital Sky Survey (SDSS) Data Release 7 \citep{York2000, Abazajian2009} with a redshift cut of $z<0.2$. In order to focus on the impact of galaxy shape and structure to morphological classifications, we utilise monochromatic $r$-band images. An extension including colour and other factors is some to consider for the future.  Here we are focused on single-band morphological classification on features seen and not in general a physical classification that might result from considering galaxy colours and colour distributions.

To examine what types of systems our classification clusters contain, as well as to have the flexibility within the data distribution in our datasets, we use morphology labels defined by T-Type \citep{deVaucouleurs1964} and the probability of being a barred galaxy ($P_{bar}$). Both quantities are obtained using deep learning techniques from  \citet[][hereafter, DS18]{DominguezSanchez2018}. We define eight labels including barred galaxies that contain significant features shown in the Hubble morphological system: ellipticals (E), lenticulars (S0), early spirals (eSp), late spirals (lSp), irregulars (Irr), barred lenticulars (SB0), early barred spirals (bar eSp), and late barred spirals (bar lSp).

The comparison of the classification scheme is shown in Table~\ref{ch6_tab:tt_hs_ourlabel}; in which, S0, eSp, and lSp are separated into barred and non-barred galaxies based on the value of $P_{bar}$. We additionally include labels of irregular galaxies from three other works: \citet{Fukugita2007}, \citet{Nair2010}, and \citet{Oh2013} to provide more irregular galaxies in our database. The morphological labels in our datasets are not used for training our machine, but to prepare an appropriate dataset with a specific data distribution, and as a way to examine the obtained clusters in terms of these types.

\begin{table}
    \begin{tabular}{|l|c|c|c|c|c|}
    \hline
    {This work} & E  & S0 & eSp & lSp & Irr \\ \hline
    {} & E  & {$S0^{-}$, S0} & S0/a - Sab & Sb - Sdm & Irr \\
    {DS18} & {-3} & {-2, -1} & {0 - 2} & {3 - 8} & 10  \\ \hline
    \end{tabular}
    \caption{The classification scheme used in this work and in \citet[][DS18; presented in T-Type]{DominguezSanchez2018}. In DS18, they define the T-Type of -3 for ellipticals ($E$), -2 for lenticulars at the early stage ($S0^{-}$), -1 for lenticulars at the intermediate to late stages ($S0$), 0 for $S0/a$, and the positive values of T-Type are for different stages of spirals. Finally the T-Type of 10 represents irregular galaxies ($Irr$).}
    \label{ch6_tab:tt_hs_ourlabel}
\end{table}

To investigate the differences in the classification systems defined by humans and those from a machine, as well as potential application within our unsupervised machine learning technique in future surveys, we prepare two different datasets: which are `balanced' and `imbalanced'. In the balanced dataset, we artificially allocate the same number of galaxy images to each morphological type. The eight human defined morphological types have visually distinctive differences from each other; therefore, the purpose of this arrangement is to allow our VQ-VAE consider fairly the characteristics of each morphology type when extracting the representative features from input images.  Otherwise it is possible that some type of bias would result if the distribution of the types we select are input into our VQ-VAE in the same fraction as they are found in the nearby universe. In this case we would find that the late-type disks would dominate over early disks and ellipticals \citep[e.g.,][]{Conselice2006}.

\begin{figure*}
\begin{center}
\graphicspath{{figures/}}
	\includegraphics[width=2.1\columnwidth]{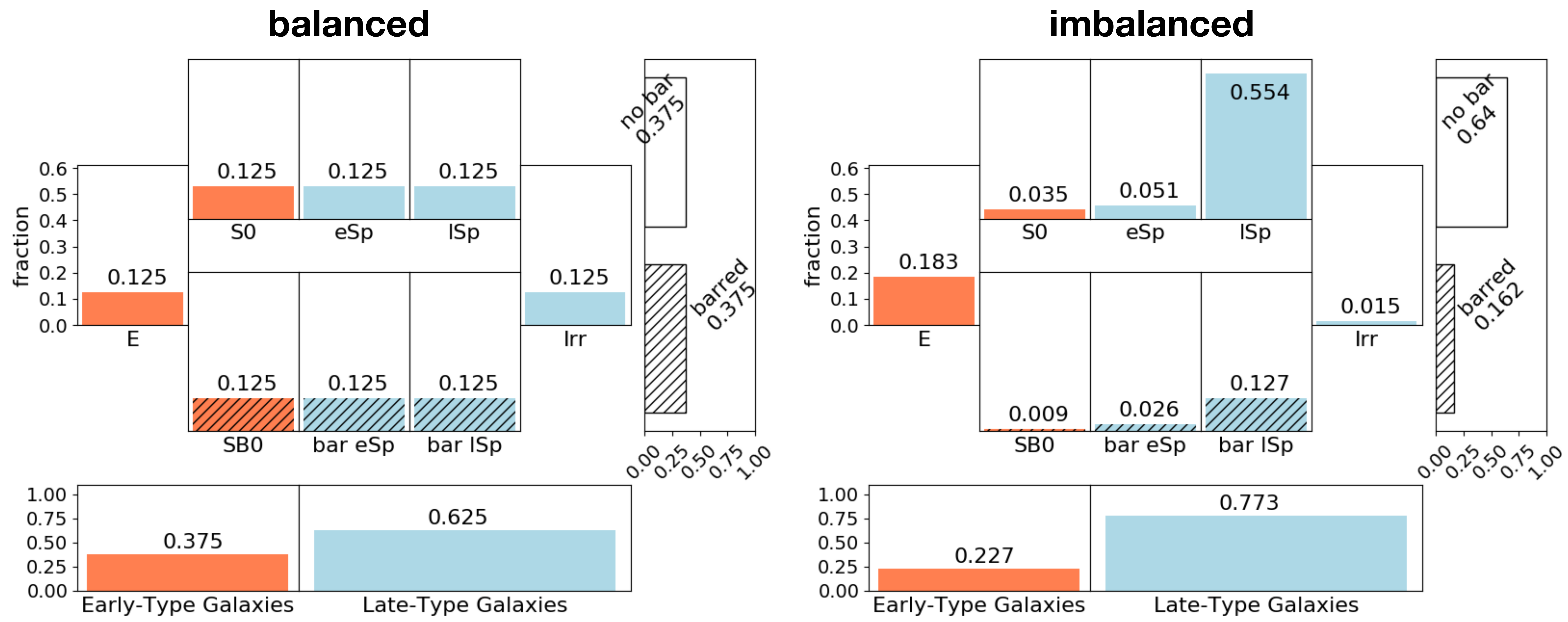}
   	\caption{The type distributions of the balanced (left) and imbalanced (right) datasets. The latter follows the distribution of nearby galaxies \citep{Oh2013}. The number shown above the coloured bar represents the fraction of the type in all data. The fraction of barred galaxies are highlighted with hashed lines. The orange and light blue colouring represent early-type galaxies and late-type galaxies, respectively.}
    \label{ch6_fig:data_type_disb}
\end{center}
\end{figure*}

On the other hand, it is of great importance to know how an unsupervised machine learning technique can be applied in future surveys to explore a large scale of unknown galaxies' morphology in an `as is' situation.  That is, we need to know how our VQ-VAE performs when galaxies are inputted from imaging observations of the real universe with no balancing.  For this goal, we set up the `imbalanced dataset' with a realistic distribution in terms of galaxy morphological types which follows the distribution of nearby galaxies at z=0.033-0.044 as presented in \citet{Oh2013}. The type distributions of the balanced and imbalanced dataset are shown in Fig.~\ref{ch6_fig:data_type_disb}.

\subsection{Feature Selection}
\label{ch6_Sec:feature_selection}

In this section we discuss a preprocessing procedure to reject irrelevant information from images. The feature selection procedure is used to select the pixels in images that are significant and which reflect the shape or structure of the targets. \citet{Cheng2020b} showed that the background noise can result in an overfit to the noise when training the convolutional autonencoder. To solve this, \citet{Cheng2020b} applied a simplified convolutional autoencoder to denoise the images and emphasise the pixels from the targets themselves before the main task is computed. However, a denoising process by another autoencoder is time-consuming and could potentially add artificial structure when reconstructing the images. Therefore, in this study, we simply use a one sigma clipping of pixel values measured through the background noises as our selection threshold. Any pixel value is below this criterion the pixel value is set as 0 \citep{Martin2020}.  Whilst this will remove noise, it will also potentially remove outer fainter portions of the galaxies themselves. However, this will retain the brighter portions of the inner parts of galaxies where classification is done in any case. Removing this fainter light does not have an effect on our measurements as it would if we were measuring for example surface brightness profiles.

\subsection{Feature Learning}
\label{ch6_Sec:feature_learning}
As described, in this study, we apply a modified vector-quantised variational autoencoder (VQ-VAE) (see Section~\ref{ch6_sec:modi_vqvae}) to carry out our unsupervised learning.  Our VQ-VAE basically learns the representative features from our images. It considers a preliminary clustering result by including an additional penalty (Equation~\ref{eq:penalty4cluster}) in the VQ-VAE (Section~\ref{ch6_sec:modi_vqvae}). This modification helps to find not only better representative features for image reconstruction, but also the features that can be well separated into two initial groups in feature space.

The main advantage of the VQ-VAE technique is to accelerate the unsupervised feature extraction process which is over 30 times faster than using a typical convolutional autoencoder \citep[e.g.,][]{Cheng2020b} without a significant trade-off to the reconstruction accuracy \citep{Razavi2019}. This is achieved by quantising the values used for reconstruction (Section~\ref{ch6_sec:vqvae}).

The hyper-parameters setting used in this study follows the setup described in \citet{Razavi2019} except for the codebook size. It determines the number of vectors available in the quantisation process (Section~\ref{ch6_sec:vqvae}). This number of vectors decides the `resolution' of the reconstructed images. Namely, the more available vectors, the more details can be presented in images. \citet{Razavi2019} use 512 vectors in their codebook to generate high-fidelity emulated images of different animals, e.g., dogs, cats. However, with a different goal from emulation in our study, we realised during analysis that a larger codebook size leads to a worse clustering result.  This is because the machine with a larger codebook uses too many details of the images into account when carrying out the clustering. These details help to complete the puzzle when emulating images but they blur the boundary in the feature space when doing clustering. In this study, after a series of heuristic tests with different codebook siezes, we choose a size of 16 for our codebook. This choice forces the machine to use the provided vectors on the most significant features and the initial two clusters show the highest silhouette score while still retaining a certain level of the reconstruction quality. This number of 16 may, and probably does, differ within different instances of use. The modified VQ-VAE models in the work trained using balanced and imbalanced datasets have a silhouette score of 0.321 and 0.306, respectively, between the two clusters obtained by K-medoids clustering.

\subsection{Clustering}
\label{ch6_Sec:clustering}

Within the clustering task, we apply an uneven iterative hierarchical clustering (Section~\ref{ch6_sec:hc}) on the data represented by a set of vector-quantised features obtained after the VQ-VAE.

In this study, we propose a new approach to decide the number of clusters within unsupervised machine learning applications. This approach can be used in other instances beyond using a VQ-VAE. Part of this is inspired by the fact that the clusters can be highly sensitive to galaxy orientation. The concept we use is to take the threshold measured by the features of galaxy orientation on the merger tree to find where the effect of galaxy orientation in a branch starts to appear (e.g., gray dotted lines in Fig.~\ref{ch6_fig:hc}). In other words, this threshold also provides the number of classification clusters that are not separated based on the galaxy orientation. This threshold is defined by the average distance between the artificially rotated images in a branch ($\overline { { d }_{ rot } }$),
\begin{equation}
\label{ch6_eq:rot_avg_pair_d}
    	\overline  { { d }_{ rot } } = \frac { \sum _{ i }^{ N }{ \sum _{ j }^{ N }{ { d }_{ ij } }  }  }{ N\left( N-1 \right)  },
\end{equation}
\noindent where $N$ is the number of datapoints in the branch, and $d_{ij}$ represents the distance between an image $i$ and image $j$. The distance, $d_{ij}$, is measured through the Hamming distance.

In this process we stop a branch and decide the number of clusters within that branch when one of two criteria is satisfied: (1) the $\overline { { d }_{ rot } }$ suggests fewer than two clusters ($\le2$) in a branch; (2) the difference between the $\overline { { d }_{ rot } }$ measured using the data of a parent branch and the data of a child branch are smaller than 0.015: that is, $\overline { { d }_{ p,rot } } -\overline { { d }_{ c,rot } } \le 0.015$.

The first criterion indicates that galaxy orientation is considered when having more than two clusters ($>2$) in this branch (e.g., circle 1 and 2 on Fig.~\ref{ch6_fig:hc}). Two clusters are the minimal number to split; therefore, we stop the iterative clustering in a branch when this criterion is satisfied. On the other hand, the second criterion is used to decide whether a branch (the parent branch) should have more sub-branches (the child branches). The variation between branches is less significant when the difference in the distance between the data of a parent branch and a child branch is small ($\le0.015$). The value used in the second criterion is measured based on the branches stopped due to the first criterion. Therefore, there is no need to split a parent branch when the second criterion is satisfied. The suggested number of clusters by the $\overline { { d }_{ rot } }$ of the parent branch is then the number of clusters in the branch without having the effect of galaxy orientation. For example in Fig.~\ref{ch6_fig:hc}, the branch stops at the circle 3 by satisfying the second criterion, and the $\overline { { d }_{ rot } }$ (gray dotted line) suggests three clusters without showing the effect of galaxy orientation in this branch.

This strategy provides a different approach for achieving ‘rotation-invariance’ in unsupervised machine learning applications. One might consider building a rotationally invariant machine learning model or to `de-rotate' galaxies as a preprocessing procedure using either mathematical methods or other machine learning techniques. For example, \citet{Martin2020} tried to produce a rotationally invariant representation using 2D Fast Fourier transforms before clustering. However, to use a rotation-invariant unsupervised machine learning model for galaxies can be time-consuming and challenging for several reasons: (1) artificially rotated galaxy images for each galaxy are essential to train a machine which enormously increases the training sample sizes; (2) it is often difficult and uncertain to determine the orientation of a galaxy; and (3) this model could be easily biased towards the training set. It is therefore difficult, uncertain, and computationally costly to pre-process and eliminate galaxy orientations with either mathematical methods or other machine learning techniques.

Therefore, in this work we propose a novel way to deal with this issue. We simply use galaxy orientation as a feature to define the clusters in such a way that we avoid generating clusters that might be sensitive to galaxy orientation. This method may unintentionally exclude galaxies with other distinctive structural features. However, the main advantage of this approach is not only to provide a different way of thinking but also to help with one of the prime issues in unsupervised machine learning applications - what is the number of clusters appropriate for a particular study?

\section{Results and Discussion}
\label{ch6_Sec:result}
\subsection{Unsupervised Binary Classification}
\label{ch6_Sec:binary_clf_ba}

\begin{figure*}
\begin{center}
\graphicspath{{figures/}}
	\includegraphics[width=2.1\columnwidth]{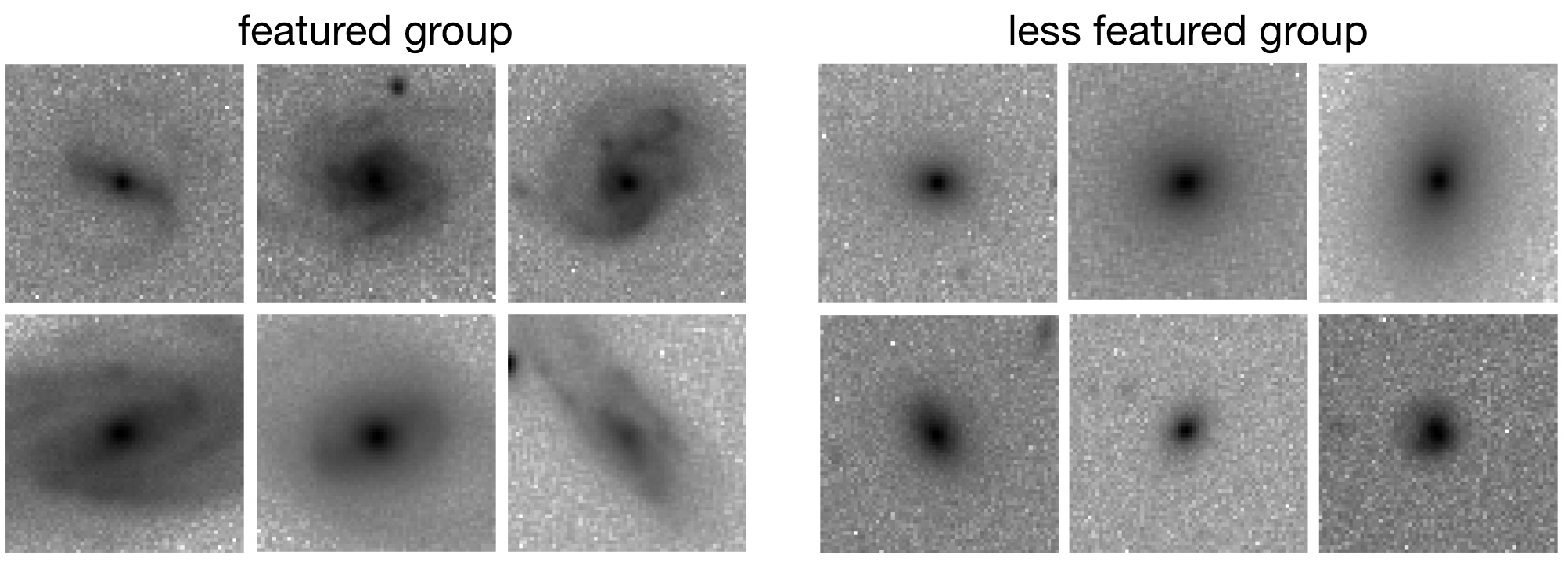}
   	\caption{Randomly picked examples of galaxies found within our two preliminary clusters using the balanced dataset. Galaxies in one cluster have more features (left), and galaxies in the other group have relatively fewer features (right).}
    \label{ch6_fig:example_ba_2cluster}
\end{center}
\end{figure*}

\begin{figure*}
\begin{center}
\graphicspath{{figures/}}
	\includegraphics[width=2.15\columnwidth]{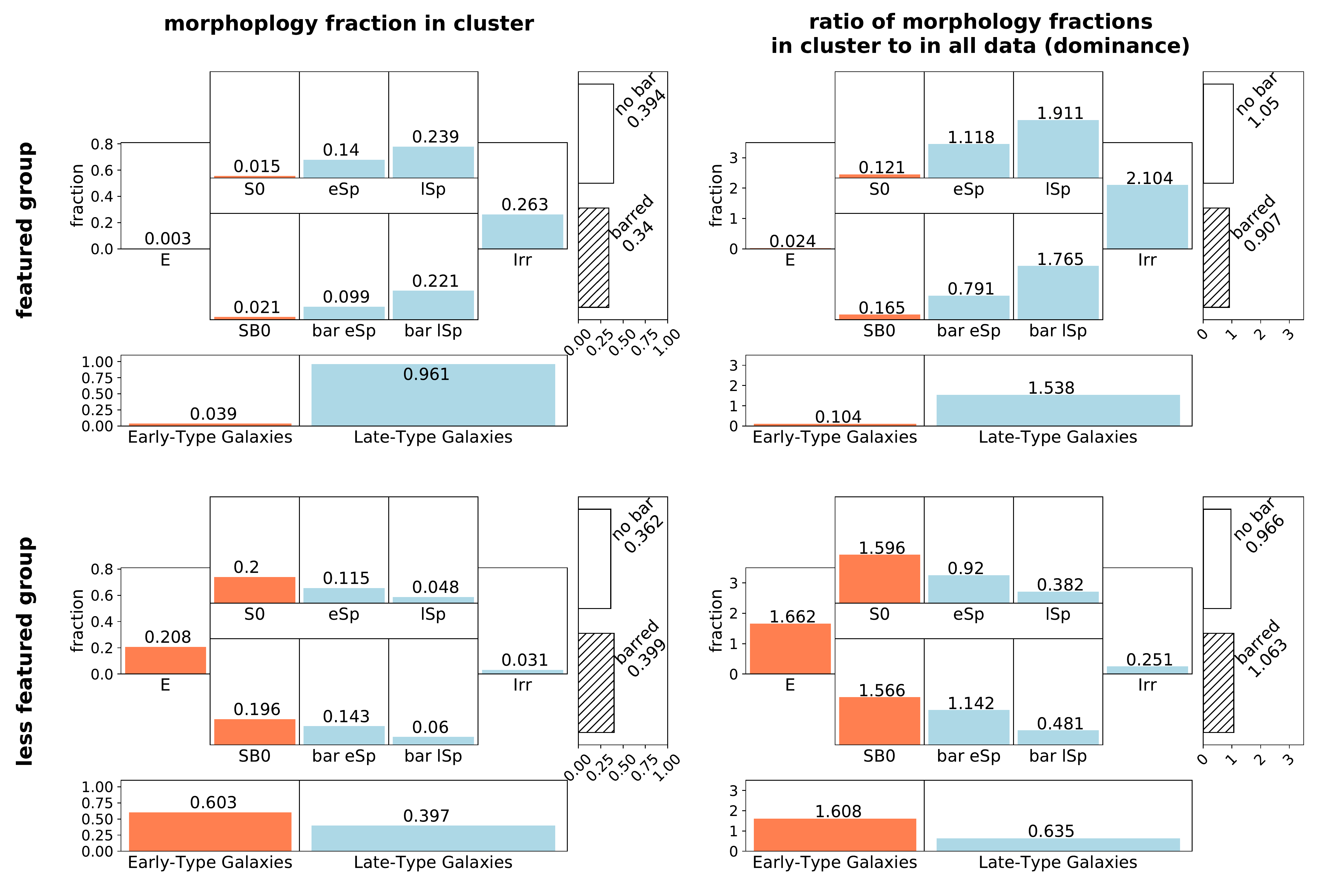}
   	\caption{The distribution of visual galaxy morphology in each cluster obtained using the balanced input dataset. The left column shows the fraction of each morphology type in the clusters while the right column presents the dominance of each type. The `dominance' is defined by the fraction of a certain morphology type in the cluster divided by a fraction of this type within the dataset. The top row shows the distribution of the `featured group' while the bottom row presents the statistics for the `less featured group'.}
    \label{ch6_fig:frac_type_ba_2cluster}
\end{center}
\end{figure*}

Starting with a simple examination, we enforce our machine to merge all galaxies in the balanced dataset into two preliminary clusters. Randomly picked examples of galaxies within the two clusters are shown in Fig.~\ref{ch6_fig:example_ba_2cluster}. Galaxies in one cluster have clearly more features (featured group; e.g., arm structure) than the galaxies of the other cluster (less featured group; more elliptical). We examine the morphological distribution in both clusters (left column in Fig.~\ref{ch6_fig:frac_type_ba_2cluster}); one cluster has $\sim96\%$ late-type galaxies (LTGs) and the other one has $\sim60\%$ early-type galaxies (ETGs).

Due to an unequal number between the ETGs and the LTGs in the balanced dataset (Fig.~\ref{ch6_fig:data_type_disb}), the fraction of ETGs and LTGs in each cluster might be biased. We examine another quantity, `dominance', which represents the ratio between the fraction of a certain type in a given cluster to the fraction of this type within the dataset (right column in Fig.~\ref{ch6_fig:frac_type_ba_2cluster}). This quantity removes the statistical influence from different number of types used in the input datasets; hence, it shows a better representation of the galaxy features emphasised in the cluster. Through the dominance distribution, we observe that the featured and less featured group are clearly dominated by the features of LTGs and ETGs, respectively.

We further investigate the potential structural factors considered when separating the two clusters. With the analysis of the two clusters, we can decide what are the major structural factors in the clustering process.  First of all it is clear that with our unsupervised learning we obtain a separation into two main clusters where one correlates with late-type galaxies and the other with early-type galaxies. This verifies with a machine this basic dichotomy which has existed in classification schemes for over 100 years.

\begin{figure*}
\begin{center}
\graphicspath{{figures/}}
	\includegraphics[width=2.1\columnwidth]{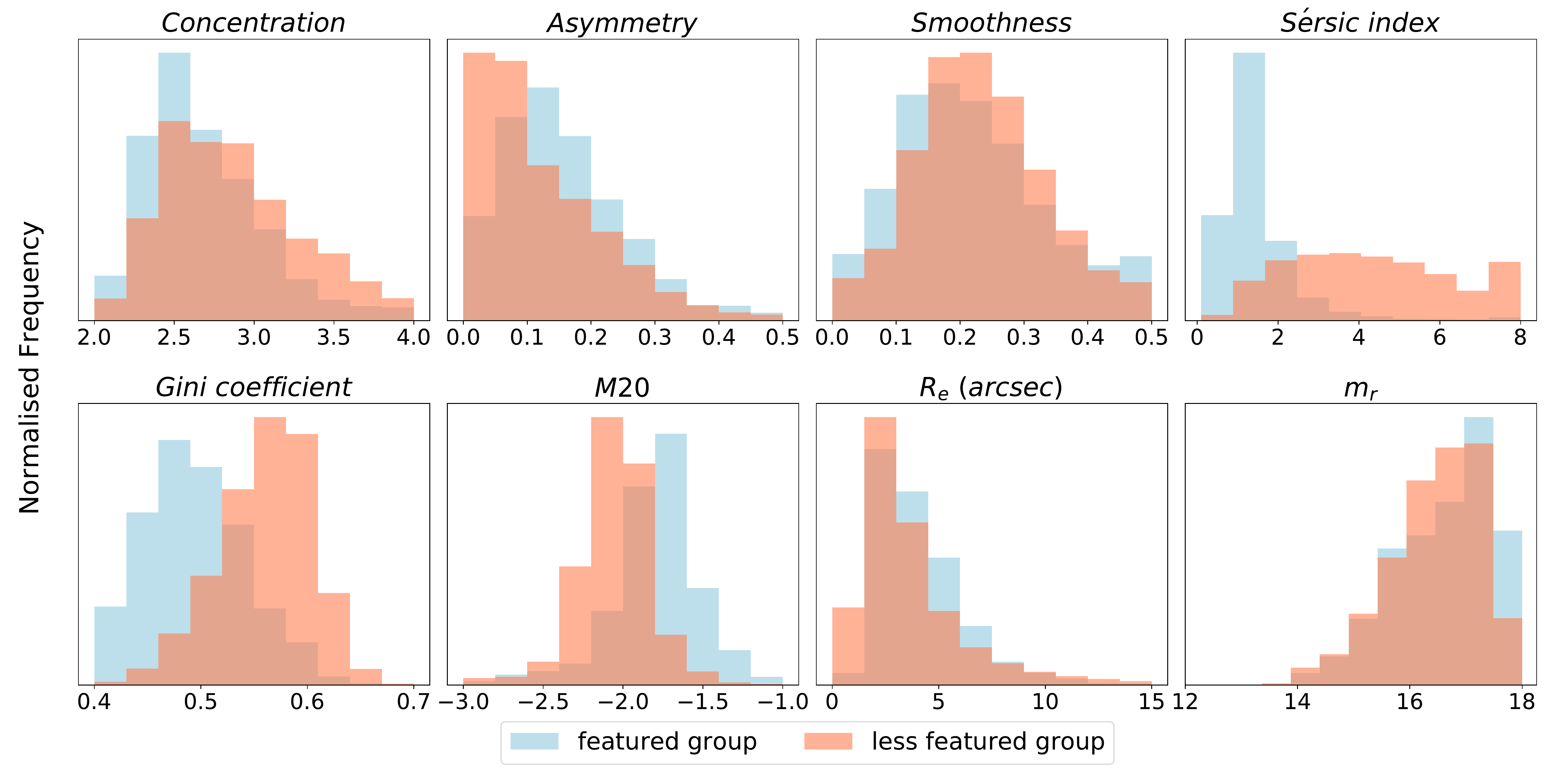}
   	\caption{The comparison of structural measurements including: concentration, asymmetry, smoothness/clumpiness, S\'ersic index, Gini coefficient, M20, half-light radius ($R_{e}$), and $r$-band apparent magnitude ($m_r$) between the two initial clusters. The blue shading represents the featured group while the orange shading is for the less featured group.}
    \label{ch6_fig:comp_structural_measurement_2cluster}
\end{center}
\end{figure*}

However, we also want to compare our clusters with more quantitiative measures.  In Fig.~\ref{ch6_fig:comp_structural_measurement_2cluster}, we compare a variety of structural measurements such as concentration, asymmetry, smoothness/clumpiness, S\'ersic index, Gini coefficient, M20, apparent half-light radius ($R_{e}$, arcsec), and $r$-band apparent magnitude ($m_r$) between the two clusters. These measurements, except for the $r$-band magnitude, are provided from the catalogue of \citet{Meert2015}, and the $r$-band magnitudes are from \citet{Simard2011}. Within these measurements, the S\'ersic index, Gini coefficient, and M20 show a clear separation, and the asymmetry shows a minor separation between the two clusters in Fig.~\ref{ch6_fig:comp_structural_measurement_2cluster}. This indicates that our machine takes galaxy structure which correlates with measurable structural parameters (asymmetry, Gini coefficient, M20) and light distribution (S\'ersic index) into account rather than the apparent size and the apparent brightness of galaxies, when categorising galaxies into the two clusters. This is good, as it shows that our method does not depend on distance or the apparent sizes of galaxies but on the inherent morphologies and structures of the galaxies themselves.

Note that the concentration and smoothness distributions show fewer differences between the two clusters. These two quantities also do not have apparent differences between the LTGs and ETGs in our dataset, because the galaxies in our datasets are relatively faint ($\sim74\%$ galaxies fainter than $m_r=16$) and the image resolution is limited by the ground-based seeing (>1 arcsec; the image sampling is 0.396 arcsec per pixel). This also produces a small separation between the two clusters in terms of asymmetry. Although we cannot straightforwardly confirm the correlation between the two clusters and the concentration parameter, the Gini coefficient and M20 provide a connection with the concept of concentration.

Based on our visual assessment, we proceed to associate the featured group to LTGs and the less featured group to ETGs in order to compare these machine-predicted labels with the catalogue labels. Using the balanced dataset, the machine-predicted and the catalogue labels agree with an accuracy of $\sim0.75$ in this binary classification. The accuracy is defined as the number of the correct matches between the machine labels and the catalogue labels from all galaxies in the dataset.

\begin{figure}
\begin{center}
\graphicspath{{figures/}}
	\includegraphics[width=\columnwidth]{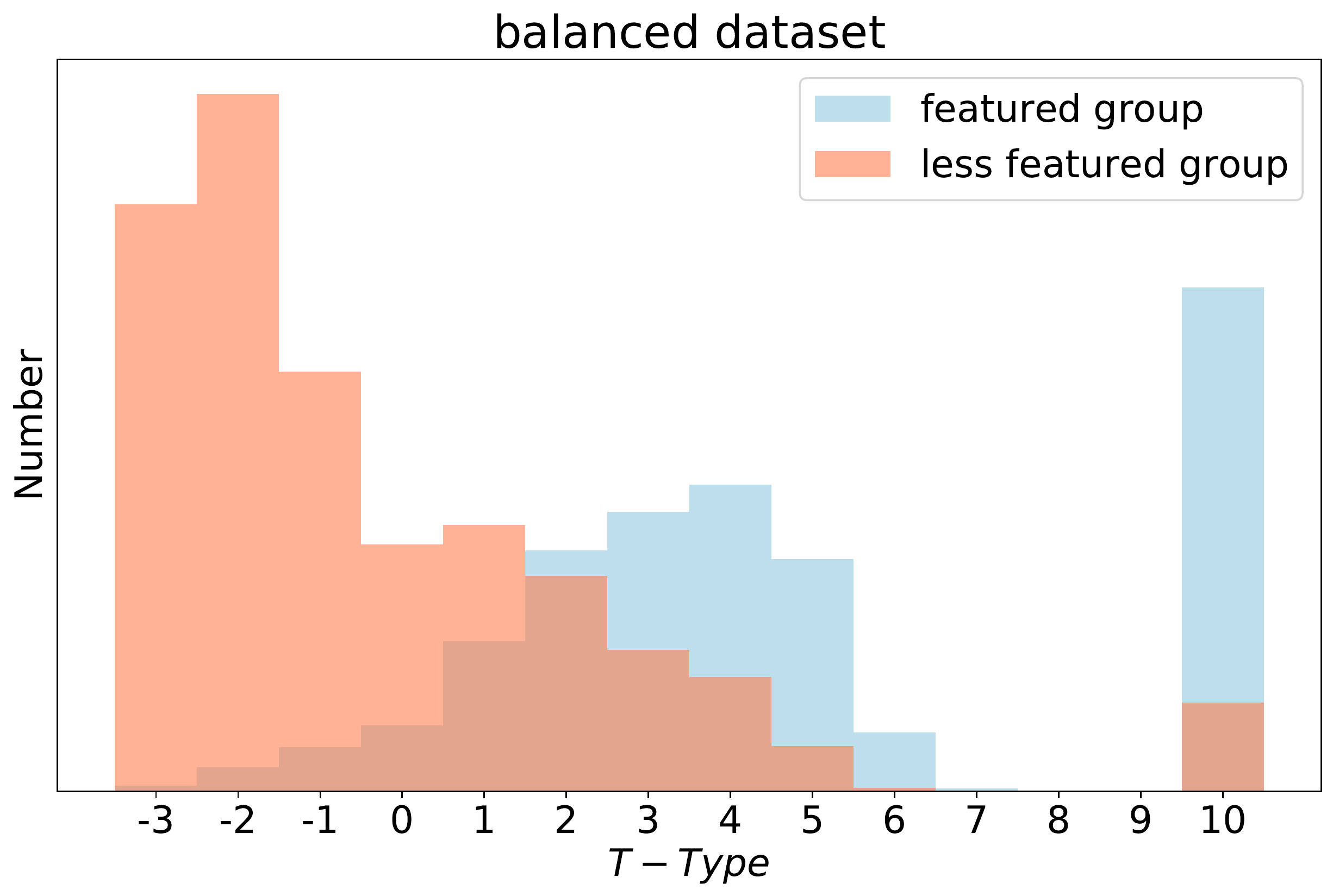}
   	\caption{The T-Type distribution between the two preliminary clusters within the balanced dataset. The corresponding visual morphology class is shown in Table~\ref{ch6_tab:tt_hs_ourlabel}. The blue shading shows the distribution of the featured group, while the light orange colour represents the less featured group.}
    \label{ch6_fig:tt_ba_2cluster}
\end{center}
\end{figure}

In Fig.~\ref{ch6_fig:tt_ba_2cluster}, we present the T-Type distribution between the two clusters. It shows that the main confusion in binary classification by our machine happens when classifying early spirals into either ETGs or LTGs, in particular, $Sab$ galaxies (T-Type=2). When we exclude early spirals from the balanced dataset, the accuracy increases to $\sim$0.87 for binary classification.

We discuss some plausible reasons for this misclassification compared to visual classification by our machine. For example, one uncertainty originates from the provided labels which combine the uncertainty of both visual classifications and machine learning predictions. Second, from our machine's perspective, in addition to the potential machine learning uncertainty, another possible uncertainty is caused by the reconstruction inaccuracy in the VQ-VAE, particularly within spiral galaxies with insignificant arm structures. However, although these causes are unavoidable, these conditions exist only in a fairly small fraction of the data in the input imaging dataset. The main reason for the mixture of early spirals in both clusters is due to the intrinsic difficulty of classifying this type into either ETGs or LTGs based only on galaxy structure. The `early spirals' in fact include a wide range of transitional features which are difficult to accurately define. The separation may become better when including colour information; however, with our method, we state the difficulty to discriminate early spirals when considering only galaxy appearance/structure in a unsupervised machine learning methodology.

\subsection{Machine Classification Scheme}
\label{ch6_Sec:machine_clf}
In the previous section, we enforce our machine to provide two initial clusters for a preliminary examination. However, the main motivation for this study is to investigate the classification system a machine would suggest when `looking' at galaxies and classifying them through machine learning. We use the proposed method in Section~\ref{ch6_sec:implementation} with the balanced dataset to let the machine explore freely and suggest a number of clusters to categorise the galaxies in the dataset. Galaxies in our dataset are categorised into 27 classification clusters by our machine. Comparing with previous work on unsupervised learning which produced 160 clusters \citep{Martin2020}. Our method suggests that significantly fewer number of galaxy morphology classifications are needed. In addition to the different implementations applied in both works, the difference in the number of obtained clusters might be due to the fact that we only consider monochromatic images to investigate the impact of galaxy structure in this study, while \citet{Martin2020} used coloured images. Additionally, to have more available measurements of galaxy structure and properties, we choose to use the imaging data from the Sloan Digital Sky Survey \citep[SDSS;][]{York2000, Abazajian2009} which has a worse resolution and image sampling (0.396 arcsec per pixel) than the one used in \citet[][0.168 arcsec per pixel]{Martin2020}. This may be a reason for the resulting fewer number of clusters obtained in our work. To further investigate galaxy morphology classifications, the colour information and images with better resolutions will be considered in future work.

\begin{figure*}
\begin{center}
\graphicspath{{figures/}}
	\includegraphics[width=2.1\columnwidth]{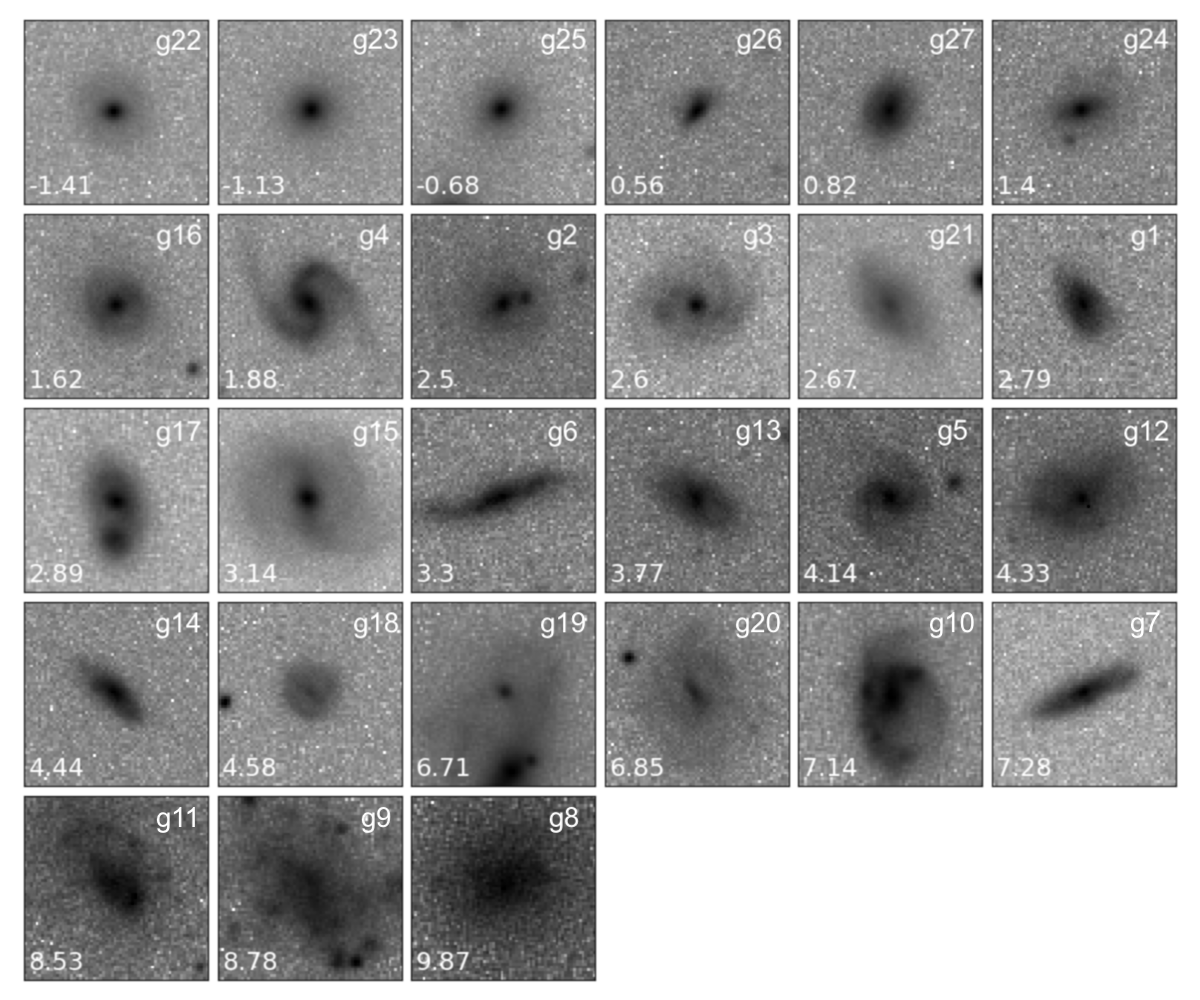}
   	\caption{Randomly picked examples of images from each cluster listed in the order of the average value of the T-Type within that cluster (Table~\ref{ch6_tab:tt_hs_ourlabel}). The number shown at the left bottom corner is the average value of the T-Type in the cluster. At the right top corner, the identification number of the belonging cluster for the image is presented.}
    \label{ch6_fig:example_ba}
\end{center}
\end{figure*}

Randomly picked examples of images from each of the 27 clusters are shown in Fig.~\ref{ch6_fig:example_ba}. The number shown on the bottom left is the average value of the T-Type in the clusters and the identification number of the cluster is shown on the top right. The identification numbers of groups are generated on the merger tree from left to right; therefore, they are simply labels without physical interpretation. Table~\ref{ch6_tab:clf_ba} lists the characteristics of each cluster in structural measurements, galaxy properties, and statistics. This can be used to co-examine the figures shown from this section to Section~\ref{ch6_Sec:machine_clf_vs_physic}. Through visual assessment in Fig.~\ref{ch6_fig:example_ba}, we find that galaxies in some clusters show bars (e.g., g15 and g16 in Fig.~\ref{ch6_fig:example_ba}, while others show more elongated shapes than others.

\begin{figure*}
\begin{center}
\graphicspath{{figures/}}
	\includegraphics[width=2.1\columnwidth]{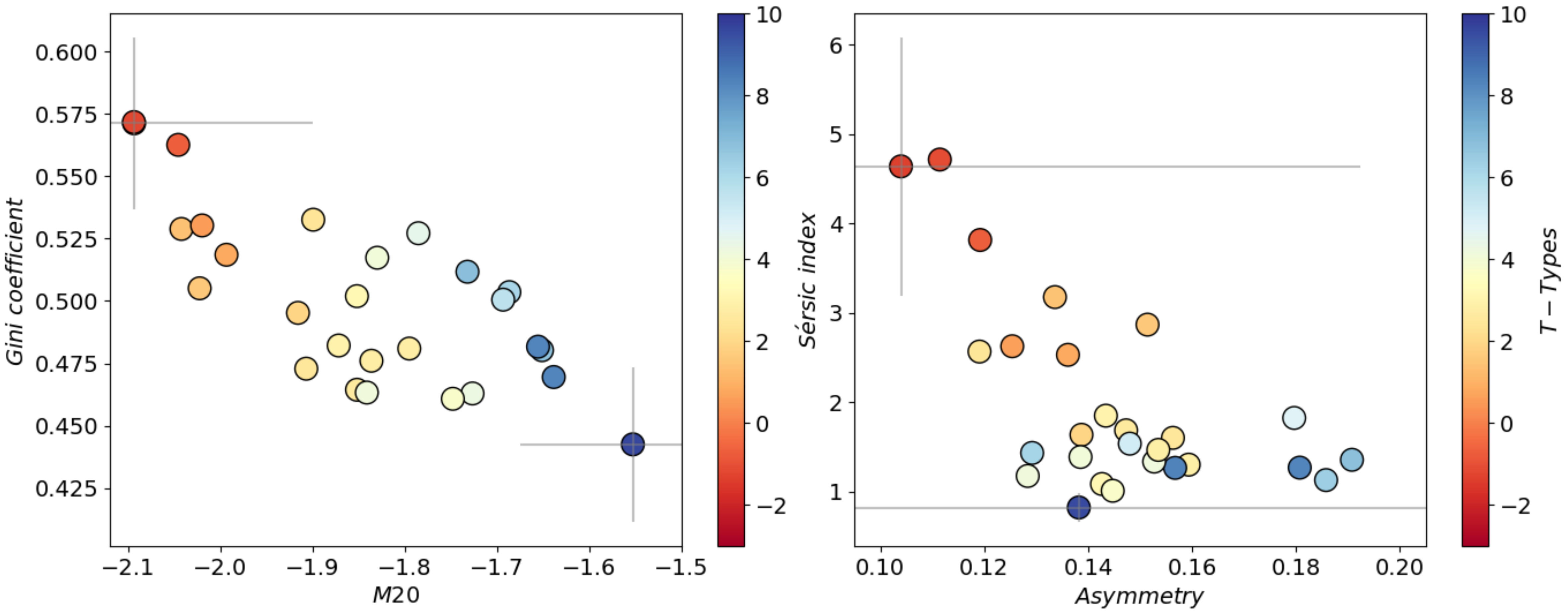}
   	\caption{The comparison of the major structural features such as the Gini coefficient, M20, S\'ersic index and the asymmetry as a function of each cluster from Section~\ref{ch6_Sec:binary_clf_ba}. Each circle represents one classification cluster from our unsupervised machine learning process which is coloured by the average T-Type in the cluster. The average value of the data in the clusters are used for each structural feature value. The gray lines show the error bars defined by the standard deviation for the two clusters with the maximum and minimum average T-Type within their clusters, respectively.}
    \label{ch6_fig:structural_feature}
\end{center}
\end{figure*}

In Fig.~\ref{ch6_fig:structural_feature}, we re-examine the influence of the major structural parameters such as the S\'ersic index, asymmetry, Gini coefficient, and M20 (Section~\ref{ch6_Sec:binary_clf_ba}), in separating clusters. Each coloured circle represents one cluster and is coloured by the average value of the T-Type in the cluster. The gray lines in Fig.~\ref{ch6_fig:structural_feature} show the error bars defined by the standard deviation for the two clusters with the maximum and minimum average values of T-Type. Asymmetry, similar to Fig.~\ref{ch6_fig:comp_structural_measurement_2cluster}, shows less discrimination between different clusters in the parameter space. Except for this, we confirm again a clear correlation between our machine classification clusters and major structural features. Additionally, the given clusters show a transition along with the T-Type. This suggests the clusters are correlated with the visual morphology roughly from early-types to late-types.

\begin{table*}
\centering
\begin{adjustbox}{width=\textwidth}
\begin{tabular}{ccccccccccccc}
  \hline
  Group & <S\'ersic n> & <Gini> & <M20> & <A> & <$g-r$> & <$Mag_{r}$> & <log$M_*$> & <$R_e$> & $N_g$ & ${ D }_{ g }$  & ${ F }_{ g,bar }$ & ${ D }_{ g,bar }$ \\
  ID & {} & {} & {} & {} & {} & {} & ($M_\odot$) & (kpc) & {(${ F }_{ g }$)} & {(${ F }_{ g,D }$)} & {}  & {(${ D }_{ g,nobar }$)} \\
  \hline
  \hline
  g1 & { 1.3 } & { 0.48 } & { -1.84 } & { 0.16 } & { 0.63 } & { -21.16 } & { 10.31 } & { 6.98 } & { 896 } & { eSp/lSp } & { 0.54 } & { 1.45 } \\
  {} & {} & {} & {} & {} & {} & {} & {} & {} & { (1.4\%) } & { (0.97) } & {} & { (1.22) } \\

  g2 & { 1.6 } & { 0.47 } & { -1.91 } & { 0.16 } & { 0.71 } & { -21.5 } & { 10.47 } & { 9.48 } & { 441 } & { eSp/lSp } & { 0.68 } & { 1.82 } \\
  {} & {} & {} & {} & {} & {} & {} & {} & {} & { (0.69\%) } & { (0.93) } & {} & { (0.83) } \\

  g3 & { 1.68 } & { 0.46 } & { -1.85 } & { 0.15 } & { 0.71 } & { -21.61 } & { 10.56 } & { 9.83 } & { 287 } & { eSp/lSp } & { 0.74 } & { 1.97 } \\
  {} & {} & {} & {} & {} & {} & {} & {} & {} & { (0.45\%) } & { (0.87) } & {} & { (0.7) } \\

  g4 & { 1.63 } & { 0.5 } & { -1.92 } & { 0.14 } & { 0.73 } & { -21.32 } & { 10.46 } & { 6.92 } & { 2924 } & { eSp/lSp } & { 0.34 } & { 0.91 } \\
  {} & {} & {} & {} & {} & {} & {} & {} & {} & { (4.57\%) } & { (0.79) } & {} & { (1.75) } \\

  g5 & { 1.17 } & { 0.46 } & { -1.84 } & { 0.13 } & { 0.52 } & { -20.19 } & { 9.79 } & { 6.52 } & { 2141 } & { lSp } & { 0.46 } & { 1.22 } \\
  {} & {} & {} & {} & {} & {} & {} & {} & {} & { (3.35\%) } & { (0.76) } & {} & { (1.3) } \\

  g6 & { 1.08 } & { 0.5 } & { -1.85 } & { 0.14 } & { 0.63 } & { -20.53 } & { 10.12 } & { 6.06 } & { 2463 } & { eSp/lSp } & { 0.14 } & { 0.37 } \\
  {} & {} & {} & {} & {} & {} & {} & {} & {} & { (3.85\%) } & { (0.8) } & {} & { (2.17) } \\

  g7 & { 1.35 } & { 0.51 } & { -1.73 } & { 0.19 } & { 0.46 } & { -20.31 } & { 9.8 } & { 5.05 } & { 3055 } & { lSp/Irr } & { 0.16 } & { 0.42 } \\
  {} & {} & {} & {} & {} & {} & {} & {} & {} & { (4.77\%) } & { (0.78) } & {} & { (0.67) } \\

  g8 & { 0.82 } & { 0.44 } & { -1.55 } & { 0.14 } & { 0.38 } & { -19.45 } & { 9.37 } & { 3.98 } & { 510 } & { Irr } & { 0.02 } & { 0.04 } \\
  {} & {} & {} & {} & {} & {} & {} & {} & {} & { (0.8\%) } & { (0.97) } & {} & { (0.03) } \\

  g9 & { 1.26 } & { 0.47 } & { -1.64 } & { 0.16 } & { 0.36 } & { -19.82 } & { 9.49 } & { 5.26 } & { 1291 } & { lSp/Irr } & { 0.16 } & { 0.43 } \\
  {} & {} & {} & {} & {} & {} & {} & {} & {} & { (2.02\%) } & { (0.94) } & {} & { (0.13) } \\

  g10 & { 1.13 } & { 0.48 } & { -1.65 } & { 0.19 } & { 0.42 } & { -20.31 } & { 9.75 } & { 5.15 } & { 946 } & { lSp/Irr } & { 0.29 } & { 0.78 } \\
  {} & {} & {} & {} & {} & {} & {} & {} & {} & { (1.48\%) } & { (0.94) } & {} & { (0.47) } \\

  g11 & { 1.27 } & { 0.48 } & { -1.66 } & { 0.18 } & { 0.36 } & { -19.88 } & { 9.49 } & { 5.2 } & { 1130 } & { lSp/Irr } & { 0.17 } & { 0.44 } \\
  {} & {} & {} & {} & {} & {} & {} & {} & {} & { (1.77\%) } & { (0.88) } & {} & { (0.29) } \\

  g12 & { 1.33 } & { 0.46 } & { -1.73 } & { 0.15 } & { 0.55 } & { -20.99 } & { 10.22 } & { 7.32 } & { 1054 } & { lSp } & { 0.74 } & { 1.99 } \\
  {} & {} & {} & {} & {} & {} & {} & {} & {} & { (1.65\%) } & { (0.85) } & {} & { (0.5) } \\

  g13 & { 1.01 } & { 0.46 } & { -1.75 } & { 0.14 } & { 0.51 } & { -20.43 } & { 9.92 } & { 6.01 } & { 941 } & { lSp } & { 0.51 } & { 1.37 } \\
  {} & {} & {} & {} & {} & {} & {} & {} & {} & { (1.47\%) } & { (0.81) } & {} & { (1.27) } \\

  g14 & { 1.39 } & { 0.52 } & { -1.83 } & { 0.14 } & { 0.63 } & { -20.62 } & { 10.16 } & { 5.7 } & { 2079 } & { eSp/lSp/Irr } & { 0.12 } & { 0.32 } \\
  {} & {} & {} & {} & {} & {} & {} & {} & {} & { (3.25\%) } & { (0.86) } & {} & { (1.76) } \\

  g15 & { 1.85 } & { 0.48 } & { -1.87 } & { 0.14 } & { 0.69 } & { -21.64 } & { 10.61 } & { 8.9 } & { 1397 } & { eSp/lSp } & { 0.73 } & { 1.94 } \\
  {} & {} & {} & {} & {} & {} & {} & {} & {} & { (2.18\%) } & { (0.87) } & {} & { (0.64) } \\

  g16 & { 2.87 } & { 0.51 } & { -2.02 } & { 0.15 } & { 0.83 } & { -22.04 } & { 10.81 } & { 11.5 } & { 776 } & { S0/eSp/lSp } & { 0.8 } & { 2.12 } \\
  {} & {} & {} & {} & {} & {} & {} & {} & {} & { (1.21\%) } & { (0.8) } & {} & { (0.51) } \\

  g17 & { 1.47 } & { 0.48 } & { -1.8 } & { 0.15 } & { 0.65 } & { -21.43 } & { 10.46 } & { 7.15 } & { 989 } & { eSp/lSp } & { 0.65 } & { 1.72 } \\
  {} & {} & {} & {} & {} & {} & {} & {} & {} & { (1.55\%) } & { (0.93) } & {} & { (0.87) } \\

  g18 & { 1.82 } & { 0.53 } & { -1.79 } & { 0.18 } & { 0.65 } & { -20.95 } & { 10.2 } & { 6.51 } & { 553 } & { eSp/lSp/Irr } & { 0.27 } & { 0.72 } \\
  {} & {} & {} & {} & {} & {} & {} & {} & {} & { (0.86\%) } & { (0.79) } & {} & { (0.98) } \\

  g19 & { 1.43 } & { 0.5 } & { -1.69 } & { 0.13 } & { 0.57 } & { -20.59 } & { 10.0 } & { 6.4 } & { 1013 } & { Irr } & { 0.17 } & { 0.46 } \\
  {} & {} & {} & {} & {} & {} & {} & {} & {} & { (1.58\%) } & { (0.59) } & {} & { (0.64) } \\

  g20 & { 1.53 } & { 0.5 } & { -1.69 } & { 0.15 } & { 0.54 } & { -20.63 } & { 9.96 } & { 6.76 } & { 982 } & { lSp/Irr } & { 0.22 } & { 0.58 } \\
  {} & {} & {} & {} & {} & {} & {} & {} & {} & { (1.53\%) } & { (0.71) } & {} & { (0.53) } \\

  g21 & { 2.56 } & { 0.53 } & { -1.9 } & { 0.12 } & { 0.76 } & { -21.29 } & { 10.46 } & { 7.8 } & { 2138 } & { S0/eSp/lSp/Irr } & { 0.29 } & { 0.76 } \\
  {} & {} & {} & {} & {} & {} & {} & {} & {} & { (3.34\%) } & { (0.68) } & {} & { (1.39) } \\

  g22 & { 4.64 } & { 0.57 } & { -2.09 } & { 0.1 } & { 0.94 } & { -22.03 } & { 10.94 } & { 7.32 } & { 12733 } & { E/S0 } & { 0.3 } & { 0.81 } \\
  {} & {} & {} & {} & {} & {} & {} & {} & {} & { (19.9\%) } & { (0.78) } & {} & { (0.87) } \\

  g23 & { 4.71 } & { 0.57 } & { -2.09 } & { 0.11 } & { 0.94 } & { -21.93 } & { 10.87 } & { 7.18 } & { 8474 } & { E/S0 } & { 0.4 } & { 1.07 } \\
  {} & {} & {} & {} & {} & {} & {} & {} & {} & { (13.24\%) } & { (0.8) } & {} & { (0.67) } \\

  g24 & { 3.17 } & { 0.53 } & { -2.04 } & { 0.13 } & { 0.81 } & { -21.82 } & { 10.73 } & { 9.14 } & { 6420 } & { S0/eSp/lSp } & { 0.69 } & { 1.85 } \\
  {} & {} & {} & {} & {} & {} & {} & {} & {} & { (10.03\%) } & { (0.69) } & {} & { (0.56) } \\

  g25 & { 3.81 } & { 0.56 } & { -2.05 } & { 0.12 } & { 0.94 } & { -21.67 } & { 10.78 } & { 6.26 } & { 3485 } & { S0 } & { 0.23 } & { 0.61 } \\
  {} & {} & {} & {} & {} & {} & {} & {} & {} & { (5.45\%) } & { (0.62) } & {} & { (1.77) } \\

  g26 & { 2.62 } & { 0.53 } & { -2.02 } & { 0.13 } & { 0.85 } & { -21.52 } & { 10.62 } & { 7.36 } & { 2056 } & { S0/eSp/lSp } & { 0.27 } & { 0.72 } \\
  {} & {} & {} & {} & {} & {} & {} & {} & {} & { (3.21\%) } & { (0.88) } & {} & { (1.89) } \\

  g27 & { 2.53 } & { 0.52 } & { -1.99 } & { 0.14 } & { 0.85 } & { -21.64 } & { 10.69 } & { 8.08 } & { 2826 } & { S0/eSp } & { 0.53 } & { 1.41 } \\
  {} & {} & {} & {} & {} & {} & {} & {} & {} & { (4.42\%) } & { (0.71) } & {} & { (1.21) } \\
\end{tabular}
\end{adjustbox}
\caption{The table lists the average values of structural measurements [S\'ersic index, Gini coefficient, M20, Asymmetry (A)] and galaxy properties [$g-r$, $r$-band absolute magnitude ($Mag_r$), stellar mass (log$M_*$), physical size ($R_e$, kpc)] in each machine-defined cluster. Additionally, the statistics of each cluster are presented in the last four columns where $N_g$ shows the number of galaxies in the cluster and $F_g$ indicates the percentage of total samples. The $D_g$ lists the dominated types in each cluster, which are selected based on the dominance of each morphology type, and $F_{g,D}$ shows the fraction of the dominated types in a cluster. The $F_{g,bar}$ is the fraction of barred galaxies in a cluster. Finally, $D_{g,bar}$ and $D_{g,nobar}$ is the dominance of barred galaxies and non-barred galaxies in a cluster, respectively. The ordering follows the group IDs which are simply labels for convenience.}
\label{ch6_tab:clf_ba}
\end{table*}

\subsection{Machine Classifications versus Human Visual Classifications}
\label{ch6_Sec:clf_vs_HS_ba}

It is important to note that the goal of this work is not to find a perfect agreement between our machine-based classification and the visual morphologies. Our goals are to understand the features used by our method to categorise galaxy images, and to introduce a novel classification scheme `proposed' by our machine.  That is, we want to develop a scheme whereby galaxies are classified by a reproducible and scientific computational way and not by human opinion.

\begin{figure*}
\begin{center}
\graphicspath{{figures/}}
	\includegraphics[width=2.1\columnwidth]{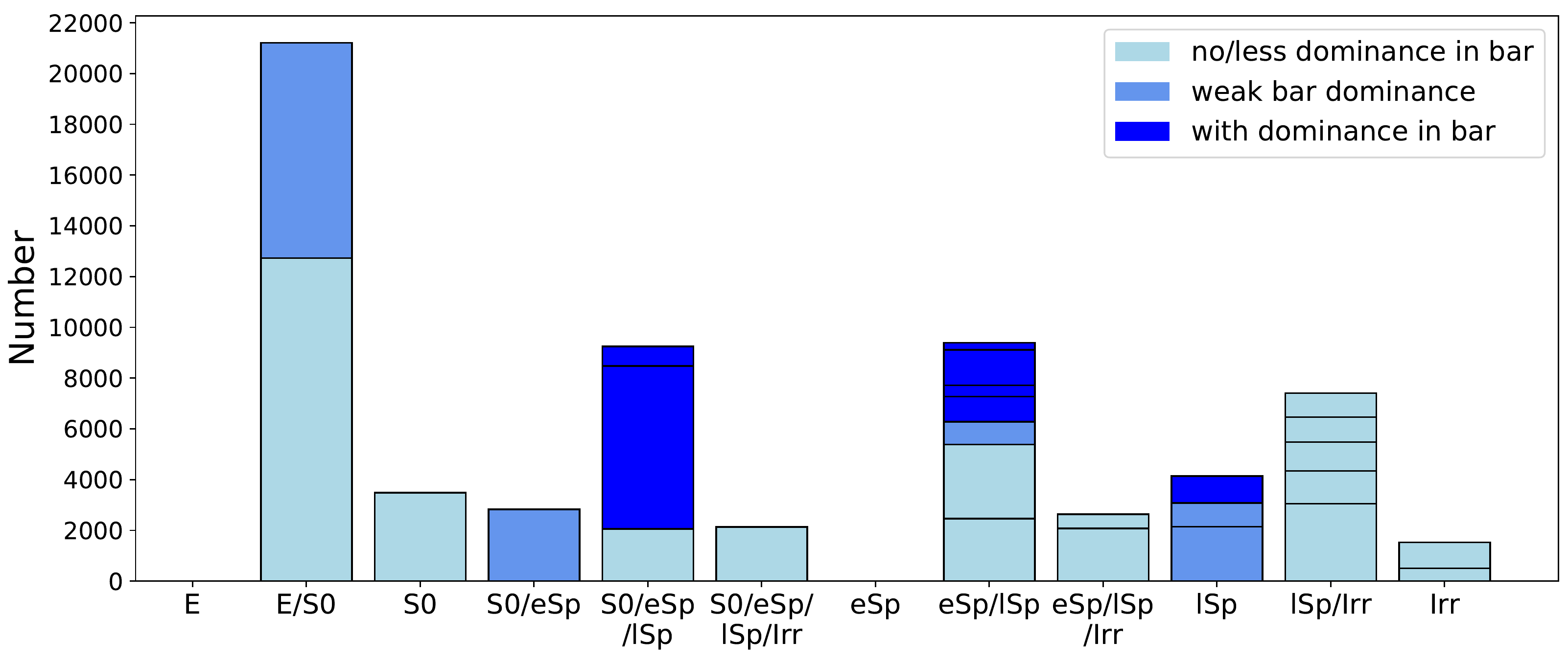}
   	\caption{The accumulated distribution of the classification clusters compared with Hubble sequence morphological types. The $x$-axis shows one or a mix of visual morphology types which dominates the clusters listed in Table~\ref{ch6_tab:clf_ba}.  All 27 clusters are plotted here, and each coloured bar represents one cluster. The different colours of the bars show different dominance levels of barred galaxies in the cluster, such that from deep to light blue represent more barred galaxies to no/fewer barred galaxies in the cluster.}
    \label{ch6_fig:ml_versus_HS_ba}
\end{center}
\end{figure*}

To better understand our machine-based classes, we compare them with visual morphological classes such as the Hubble sequence, and discuss the visual features extracted by our machine. To do this comparison, we associate each cluster with one or a mix of Hubble types based on the dominance of each type within each of the clusters (Fig.~\ref{ch6_fig:ml_versus_HS_ba}). As mentioned in Section~\ref{ch6_Sec:binary_clf_ba}, the `dominance' of each type is the ratio between the fraction of a given morphology type in the cluster to the fraction in the dataset. We associate a given cluster with one or several morphology types when the dominance of a certain type is $>1$. This selection indicates which kinds of visual features considered in a visual morphology type are dominated in a cluster.

\begin{figure*}
\begin{center}
\graphicspath{{figures/}}
	\includegraphics[width=2.1\columnwidth]{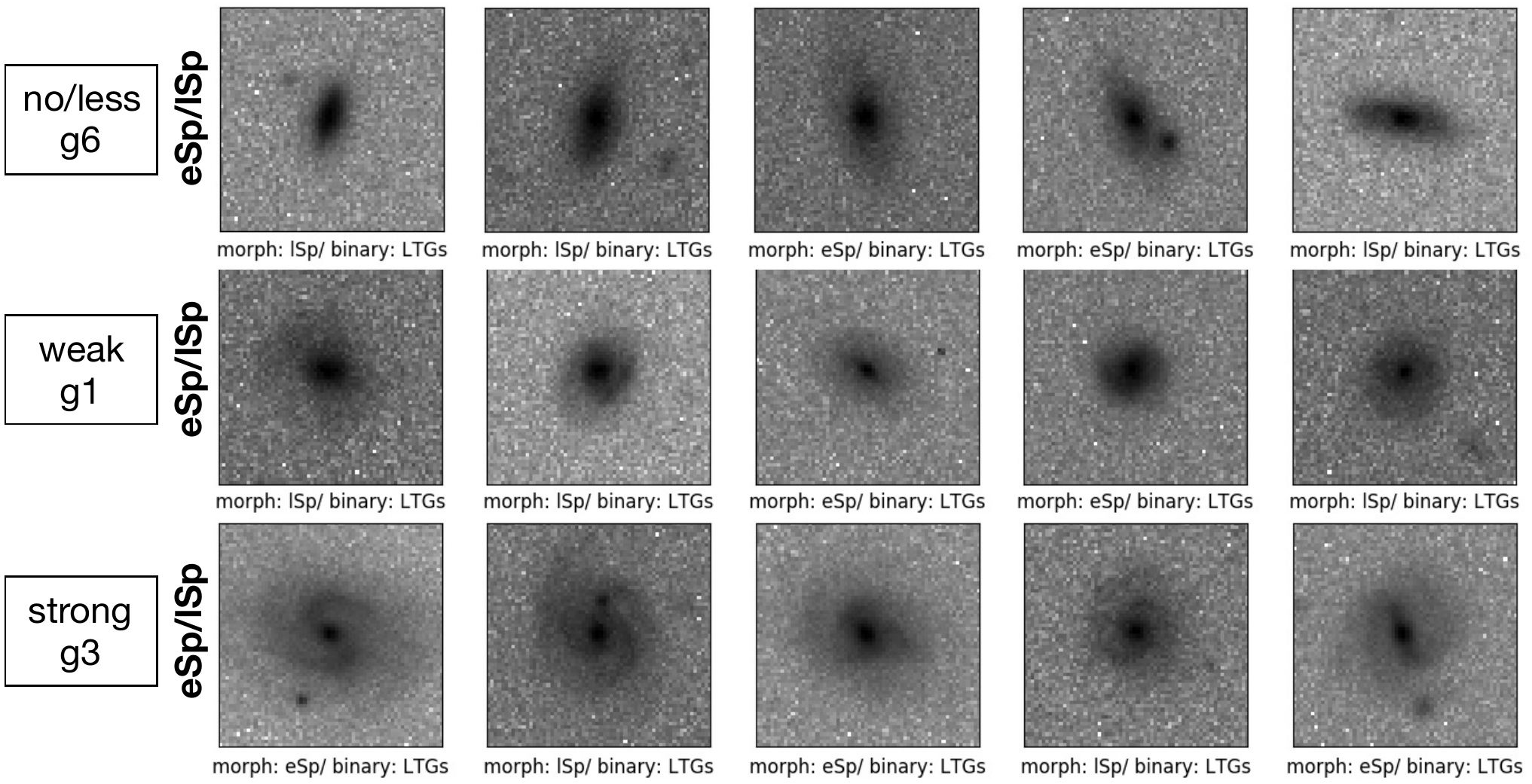}
   	\caption{Examples of the clusters with different bar dominance levels. Each row shows five randomly picked examples in the cluster, where `g6' represents the 6th cluster, and so on. From top to bottom, examples of no/less, weak, strong bar dominance are presented, respectively. The galaxy morphology information is shown below each image.}
    \label{ch6_fig:example_different_bar}
\end{center}
\end{figure*}

In Fig.~\ref{ch6_fig:ml_versus_HS_ba}, we show the accumulated distribution of the classification clusters to one or a mix of visual morphology types. Each coloured bar represents one cluster and the deeper bluer colours indicate more barred galaxies than non-barred galaxies within that given cluster. In Fig.~\ref{ch6_fig:ml_versus_HS_ba}, the darkest blue represents a cluster with the strong bar dominance, $D_{g,bar}\ge1$ and the non-bar dominance, $D_{g,nobar}<1$ (see the last column in Table~\ref{ch6_tab:clf_ba}; e.g., g16 in the table). The medium blue is for a cluster with both bar and non-bar dominance $\ge1$ (weak bar dominance; e.g., g27 in Table~\ref{ch6_tab:clf_ba}). This criterion indicates that the features of a barred galaxy are not distinctive in a cluster. The lightest blue is used when the bar dominance is $D_{g,bar}<1$ (no/less dominance; e.g., g14 and g19 in Table~\ref{ch6_tab:clf_ba}). Through the highlight of the bar dominance in clusters in Fig.~\ref{ch6_fig:ml_versus_HS_ba}, our machine is shown to successfully discriminate between barred and non-barred galaxies. Examples of clusters with different bar dominance are shown in Fig.~\ref{ch6_fig:example_different_bar}.

\begin{figure}
\begin{center}
\graphicspath{{figures/}}
	\includegraphics[width=\columnwidth]{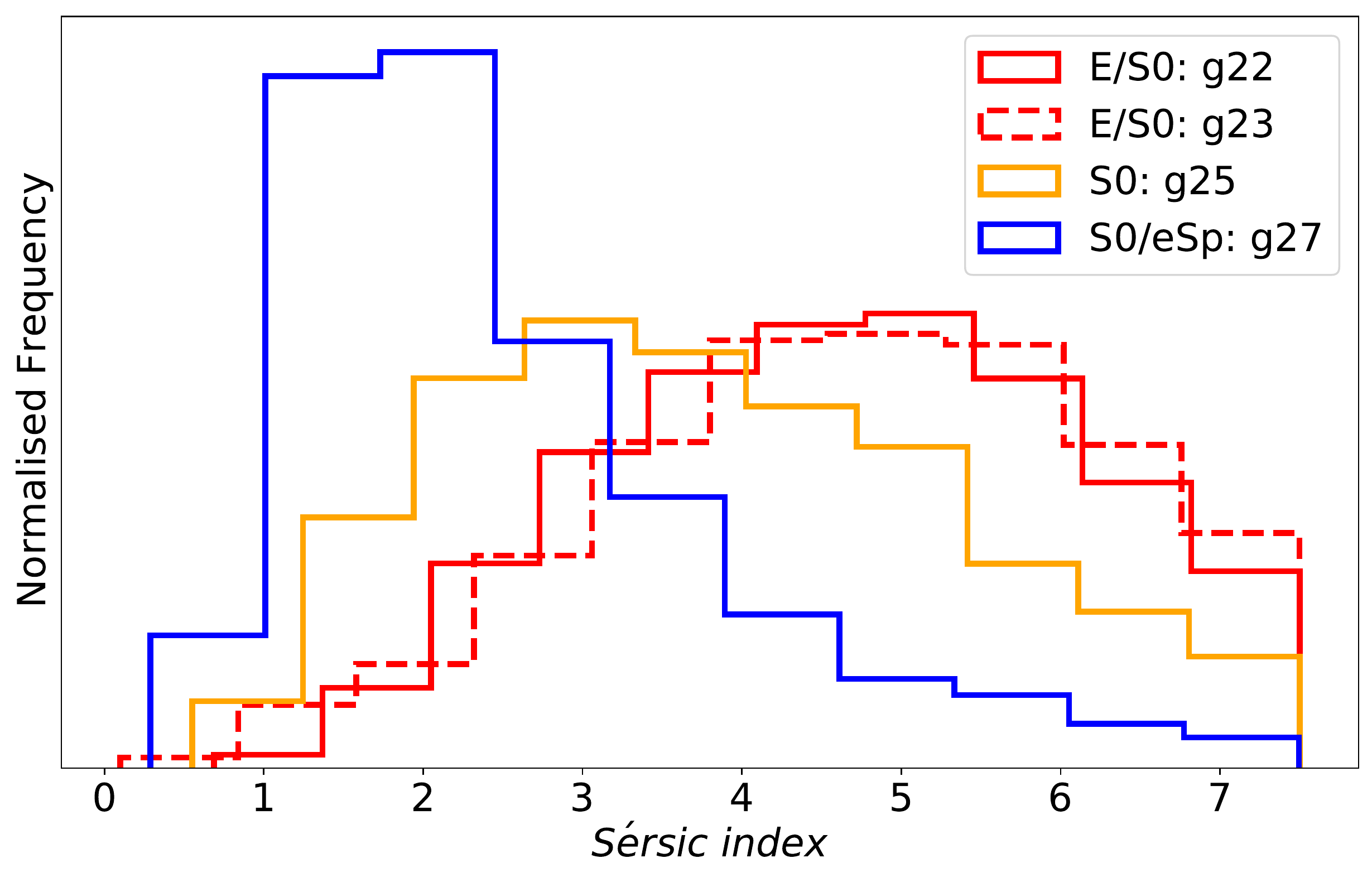}
   	\caption{The S\'ersic index distribution for the clusters dominated by $E/S0$ galaxies (g22: red solid line; g23: red dashed line), $S0$ (g25: yellow solid line), and $S0/eSp$ (g27 : blue solid line), where `g22' represents the 22th cluster, and so on.}
    \label{ch6_fig:comp_sersic_E_S0_esp}
\end{center}
\end{figure}

We observe in Fig.~\ref{ch6_fig:ml_versus_HS_ba} that no cluster is dominated by either elliptical galaxies or early spirals only. The features of elliptical galaxies are recognised to have a great similarity to some lenticular galaxies by our machine. Visually, we separate ellipticals and lenticulars mainly based on the disk structure. However, compared to the cluster dominated by only lenticulars (the g25 in Table~\ref{ch6_tab:clf_ba}) in Fig.~\ref{ch6_fig:comp_sersic_E_S0_esp}, the galaxies in the two clusters dominated by $E/S0$ (g22; g23) lack significant disk structure, whereas `g22' represents the 22th cluster, and so on (also see Fig.~\ref{ch6_fig:example_ba} and Table~\ref{ch6_tab:clf_ba}). However, clusters with more disky galaxies, such as g27 (blue solid line in Fig.~\ref{ch6_fig:comp_sersic_E_S0_esp}), are dominated by a mix of $S0$ and $eSp$. This is likely an indication for an uncertainty in distinguishing ellipticals, lenticulars, and early spirals in the visual classification system we use and not a defect of our unsupervised learning. Only the lenticulars with a moderate range of S\'ersic index (peaks at $\sim3$; yellow solid line in Fig.~\ref{ch6_fig:comp_sersic_E_S0_esp}) can be separated from other morphology types.

Additionally, as stated in Section~\ref{ch6_Sec:binary_clf_ba}, early spirals are difficult to categorised into either ETGs or LTGs, and as such it is difficult to have a distinctive cluster dominated by only this morphology type (Fig.~\ref{ch6_fig:ml_versus_HS_ba}) due to the broad transitional features in this type. This again indicates the intrinsic difficulty of visually separating early spirals from other morphology types, such as lenticulars and late spirals.

\begin{figure*}
\begin{center}
\graphicspath{{figures/}}
	\includegraphics[width=2.1\columnwidth]{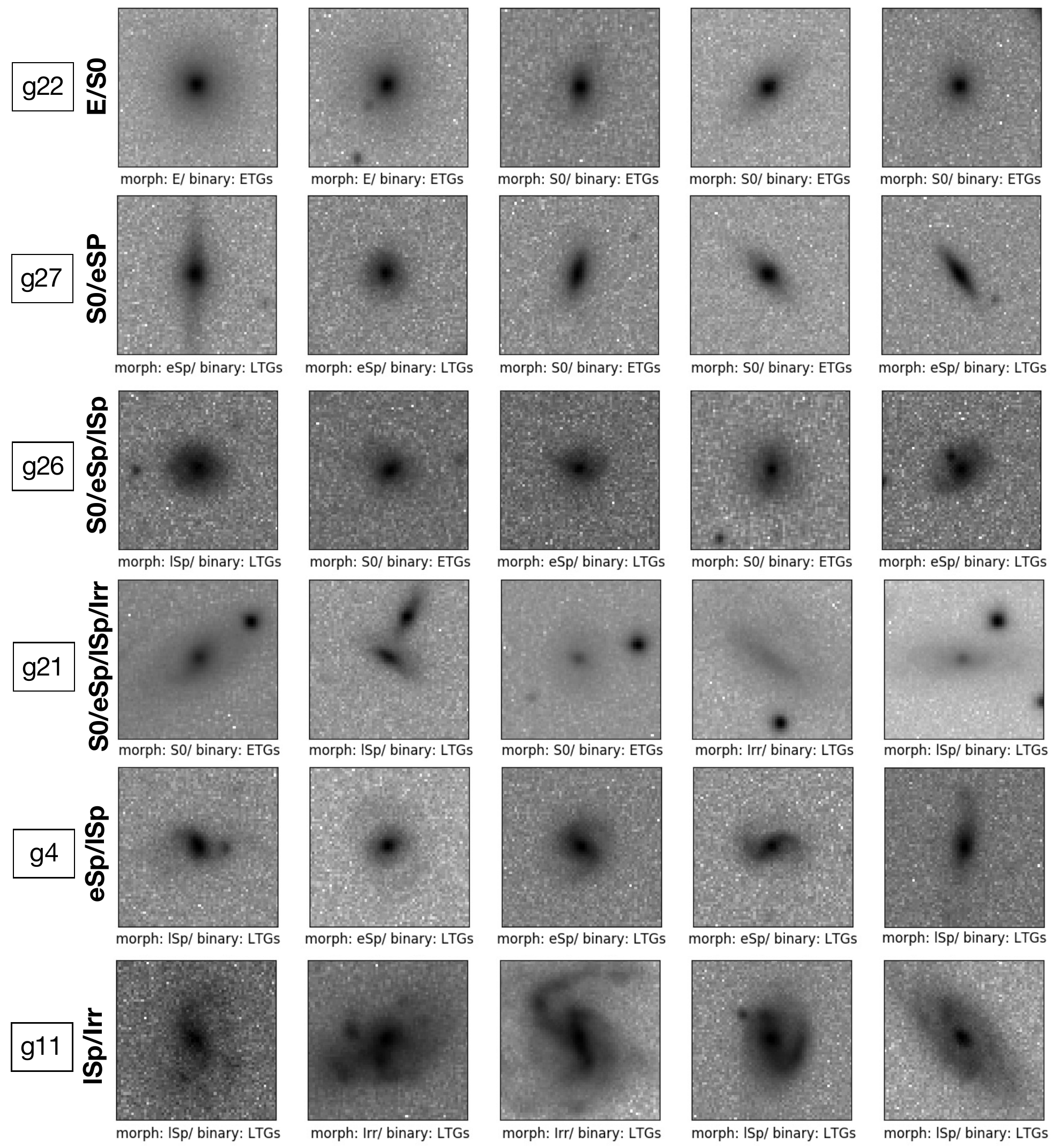}
   	\caption{Examples of images of galaxies from clusters with a mix of many visual morphology types. Each row shows five randomly picked examples within the cluster, where `g22' represents the 22th cluster, and so on. The morphology information is shown below each image.}
    \label{ch6_fig:example_mixture_types}
\end{center}
\end{figure*}

Most of our clusters have a mixture of different Hubble types within them which indicates galaxies with similar features in appearance can be visually classifying into a variety of morphology types (see examples in Fig.~\ref{ch6_fig:example_mixture_types}). In other words, a mix of galaxy structure in fact exists in a visually defined morphology type. This result reveals an intrinsic vagueness of the visual classification systems such that they are not always accurately defined, with many galaxies not optimally classified as a certain T-Type due to the diversity of properties beyond a guessed at morphology.

One exception from the above discussion is our cluster 21 (g21 in Table~\ref{ch6_tab:clf_ba} with a mix of four morphology types: S0, eSp, lSp, Irr). This cluster is shown to have galaxies with bright companions which overwhelms the brightness of the central objects (the `g21' row shown in Fig.~\ref{ch6_fig:example_mixture_types}). After the feature selection and normalisation in Section~\ref{ch6_Sec:feature_selection}, the central objects might become negligible to the machine learning compared to the companions. This can result in  difficulty for our machine to capture the structure of the central objects as well as group these galaxies correctly. On the other hand, galaxies with companions are more likely to experience galaxy mergers, and thus this cluster can be used as an indication to find potential merger events or compact groups of galaxies.






\subsection{Machine Classifications versus Physical Properties}
\label{ch6_Sec:machine_clf_vs_physic}

In previous sections, we show that our machine learning classifications trained with monochromatic images are categorised based on structural features (Section~\ref{ch6_Sec:machine_clf}) and visual features (Section~\ref{ch6_Sec:clf_vs_HS_ba}). In this section, we examine several galaxy properties in each machine-defined galaxy class.

\begin{figure*}
\begin{center}
\graphicspath{{figures/}}
	\includegraphics[width=2.1\columnwidth]{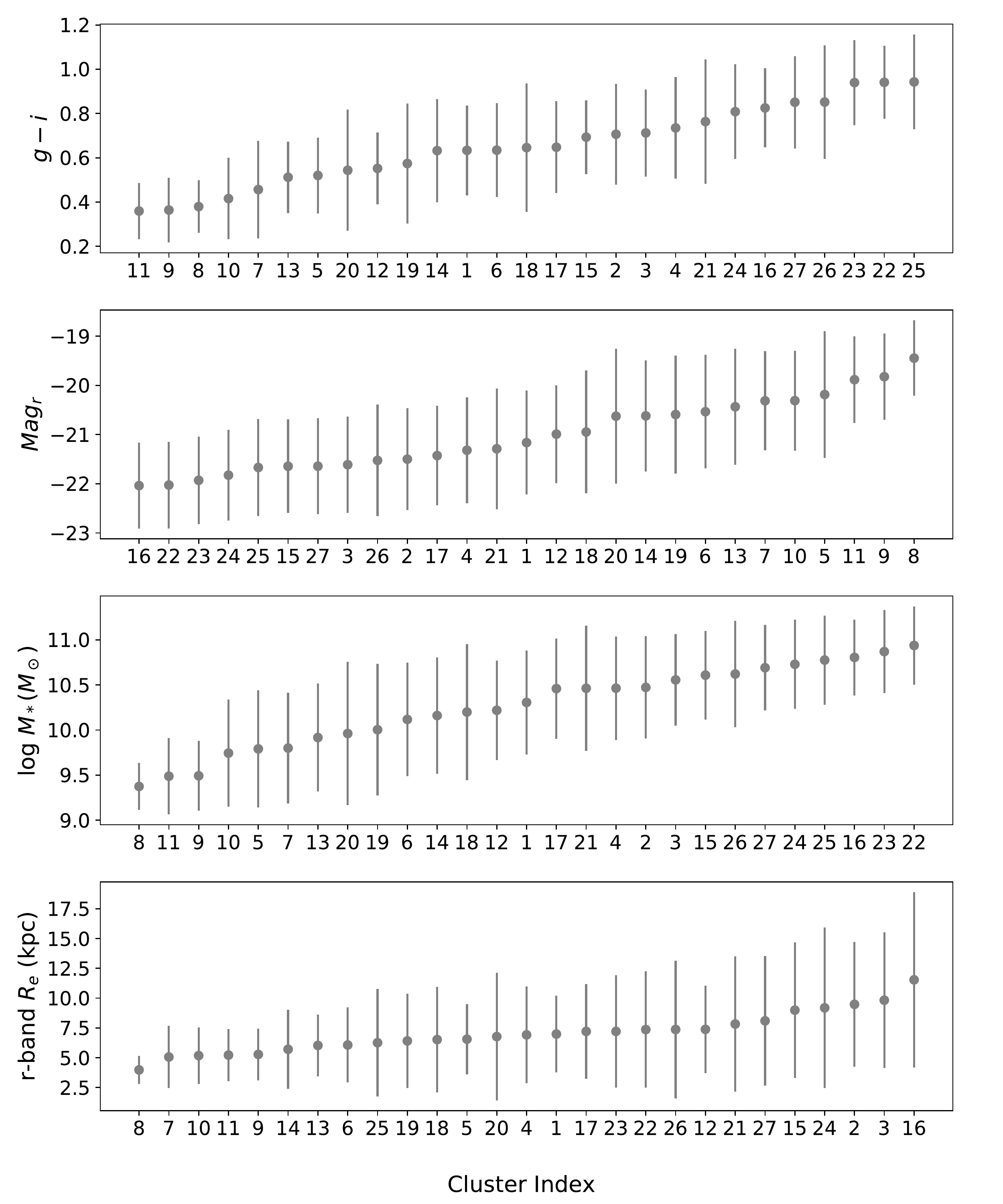}
   	\caption{The diagram shows the average value of different galaxy properties such as $g-i$, $r$-band absolute magnitude ($Mag_{r}$), stellar mass ($M_*$), and galaxy physical sizes ($R_e$, kpc) for each machine-defined cluster. The error bars show the standard deviation of the values within a cluster. The $x$-axis represents the cluster index without any physical meaning. Each individual plot is ordered by the average value of each physical property for the galaxies in the clusters.}
    \label{ch6_fig:cluster_vs_physical_properties}
\end{center}
\end{figure*}

First, Fig.~\ref{ch6_fig:cluster_vs_physical_properties} shows the average values of different galaxy properties such as $g-i$ colour, $r$-band absolute magnitude ($Mag_{r}$), stellar mass ($M_*$), and galaxy physical sizes ($R_e$, in kpc) for each machine-defined cluster. The colours and physical sizes are again taken from \citet{Simard2011} while the stellar mass originates from \citet{Mendel2014}. Each cluster, as defined by the machine in this plot, has distinctive physical properties in galaxy colour, absolute magnitude, stellar mass, and physical size. Even though the discrimination can be small in value when considering the error bars, a clear transition of physical properties along with different clusters are shown.

\begin{figure*}
\begin{center}
\graphicspath{{figures/}}
	\includegraphics[width=2.1\columnwidth]{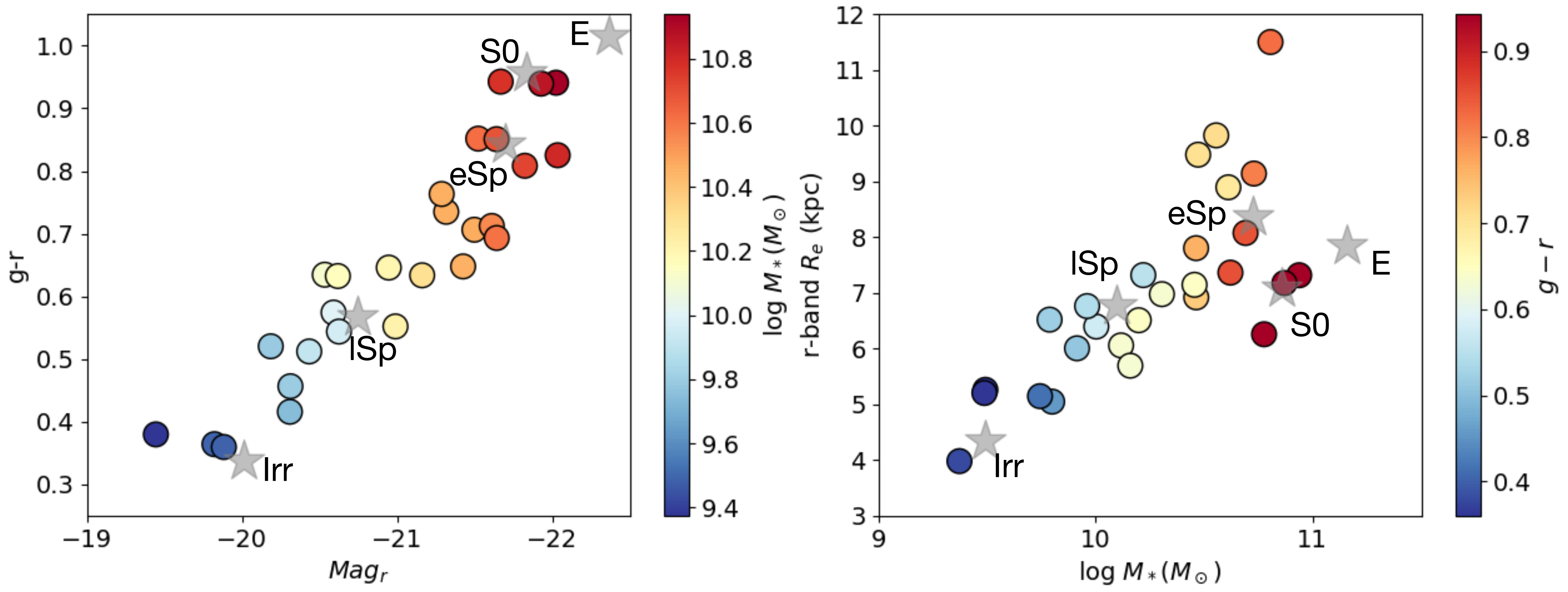}
   	\caption{{\it Left:} the colour-magnitude diagram of the classification clusters where the $x$-axis is the average values of the $r$-band absolute magnitude ($Mag_{r}$) and the $y$-axis represents the average value of the galaxy colours ($g-r$) within each plotted cluster. Each circle represents one classification cluster from our unsupervised machine and coloured by the average value of the stellar mass ($M_*$). {\it Right:} the mass-size relation of the given clusters where the $x$-axis and $y$-axis is the average values of the stellar mass ($M_*$) and the average values of the galaxy physical sizes ($R_e$, kpc), respectively. Each circle is coloured by the average value of galaxy colour ($g-r$). In both graphs, each star shows the average values of these quantities for the traditional Hubble types for comparison, where the type of each is written in black.}
    \label{ch6_fig:comp_physical_properties}
\end{center}
\end{figure*}

In Fig.~\ref{ch6_fig:comp_physical_properties} we combine these properties together and plot the colour-magnitude plane (left) and the mass-size plane (right). Each circle represents one cluster, coloured by the average value of the stellar mass of the galaxies in the cluster for the colour-magnitude diagram, and by the average colour value for the mass-size relations. Each star shows the average value of the data within a certain visual morphology type (written in black) for comparison. The machine-defined morphology types fill in the gap within the correlation of galaxy morphology and galaxy properties along with the Hubble types. This indicates that the machine classification scheme can construct the missing morphologies in the visual classification systems without involving human potential bias. It will be interesting to investigate the correlation of these machine-defined classifications with galaxy environment and other galaxy properties, which will be left to study in a future paper.

Additionally, we notice on the mass-size diagram (right in Fig.~\ref{ch6_fig:comp_physical_properties}) that the five orange clusters above the {\it eSp} star-label are dominated by barred galaxies, in particular, the top cluster with the largest average size has $\sim80\%$ barred galaxies in the cluster (g16 in Table~\ref{ch6_tab:clf_ba}). Galaxies in this cluster have larger sizes, larger stellar masses, and are redder in colour than other clusters with a mix of typical spiral galaxies.

\subsection{Dataset with a real distribution}
\label{ch6_Sec:im_test}

To test the capability of our method on a realistic data distribution, we apply our method to the imbalanced dataset (Fig.~\ref{ch6_fig:data_type_disb}) which follows the distribution of intrinsic morphology for nearby galaxies \citep[][Section~\ref{ch6_sec:datasets}]{Oh2013}. In this section, we examine the performance using this dataset for: (1) binary classification (Section~\ref{ch6_Sec:binary_clf_im}) and (2) multiple classification clusters (Section~\ref{ch6_Sec:ml_clf_im}) using the imbalanced dataset, and compare the results with the one using the balanced dataset.

\subsubsection{Unsupervised binary classification}
\label{ch6_Sec:binary_clf_im}

\begin{figure*}
\begin{center}
\graphicspath{{figures/}}
	\includegraphics[width=2.1\columnwidth]{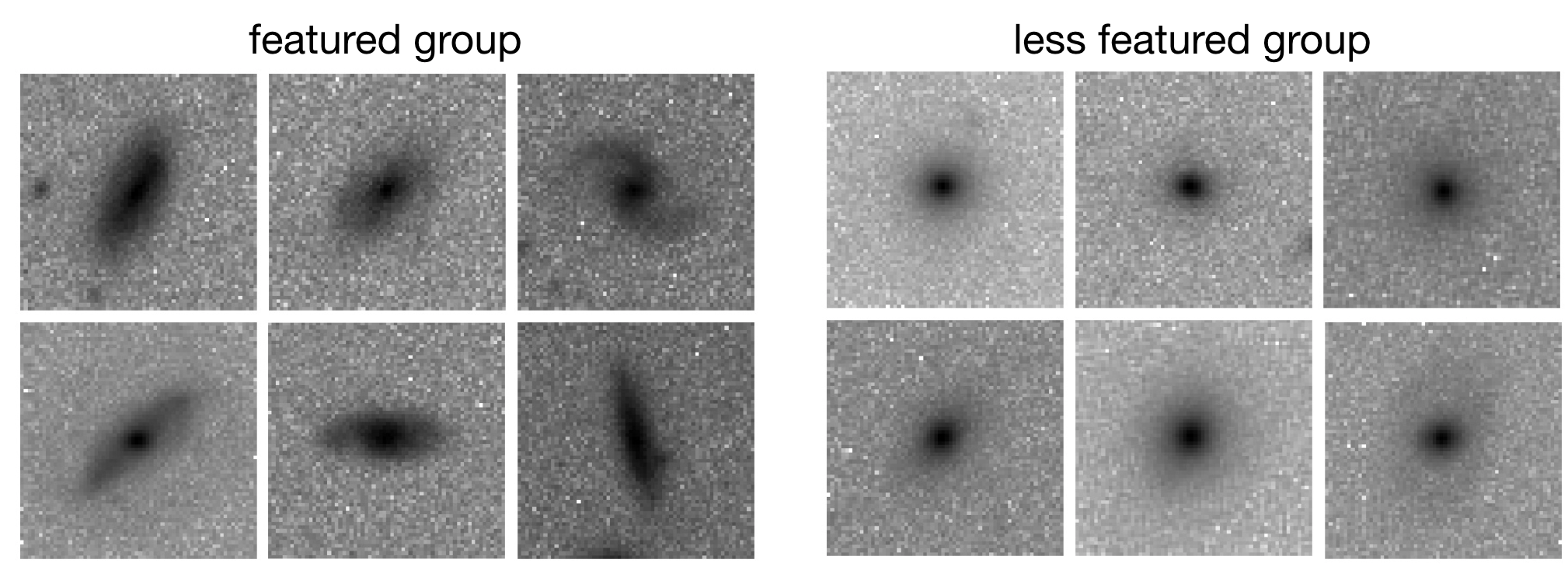}
   	\caption{Examples of galaxies within the  two preliminary clusters using the imbalanced dataset. Galaxies in one cluster are with more features (left), and galaxies in the other group are with relatively fewer features (right).}
    \label{ch6_fig:example_im_2cluster}
\end{center}
\end{figure*}

Similar to Section~\ref{ch6_Sec:binary_clf_ba} for the balanced dataset, we merge the imbalanced dataset into two preliminary clusters (Example of galaxies in each is shown in Fig.~\ref{ch6_fig:example_im_2cluster}). Although the imbalanced data has a significantly different distribution in galaxy types from the balanced dataset, our machine obtains two preliminary clusters with similar features to the two clusters provided using the balanced dataset (Fig.~\ref{ch6_fig:example_ba_2cluster}). As before, one cluster is dominated by galaxies with many distinct features while the other has galaxies with significantly fewer features.

\begin{figure*}
\begin{center}
\graphicspath{{figures/}}
	\includegraphics[width=2.15\columnwidth]{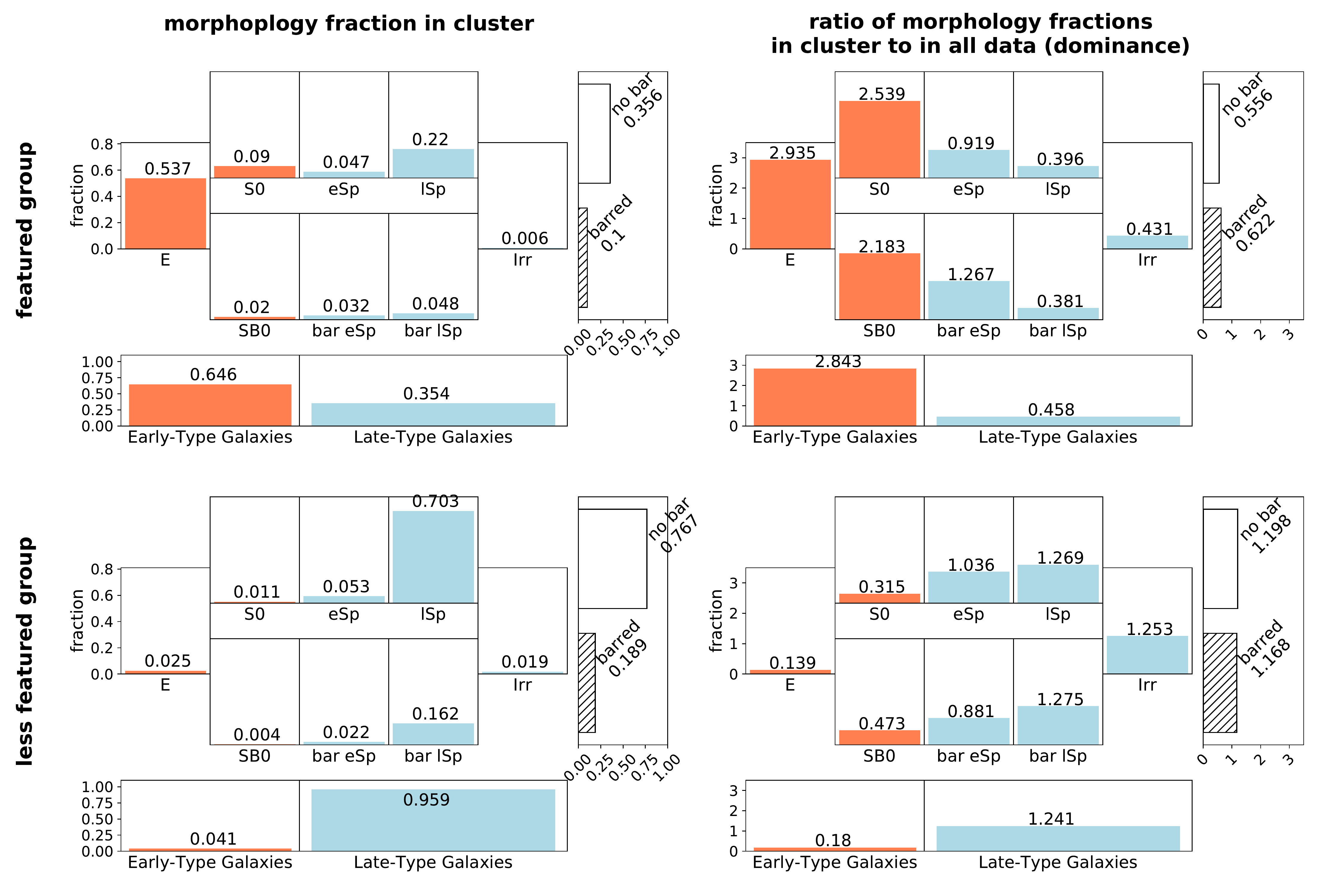}
   	\caption{The distribution of visual galaxy morphology in each cluster obtained using the imbalanced dataset. The left column shows the fraction of each morphology type in the clusters, while the right column shows the dominance of each type. The top row shows the distribution of the `featured group' while the bottom row presents the one of the `less featured group'.  This can be compared to the same distribution when using the balanced dataset shown in Fig.~\ref{ch6_fig:frac_type_ba_2cluster}.}
    \label{ch6_fig:frac_type_im_2cluster}
\end{center}
\end{figure*}

Fig.~\ref{ch6_fig:frac_type_im_2cluster} shows the morphological fractions of different types (left column) and the dominance of each morphology type in each cluster (right column). The dominance is, again,  the ratio between the morphological fraction in the cluster to the fraction in the dataset. This quantity removes the impact of the imbalanced numbers between each type, and indicates the visual features emphasised in a cluster. The two clusters are clearly dominated by LTGs and ETGs, respectively. Additionally, the dominance distribution of the imbalanced dataset is completely consistent with that of the balanced dataset (Fig.~\ref{ch6_fig:frac_type_ba_2cluster}). This confirms that no matter which data distribution is used, our machine is capable of separating the two clusters based on the specific features existing in the corresponding morphology types.

Additionally, applying our method to the imbalanced dataset we get an initial accuracy of $\sim$0.87 in separating ETGs from LTGs. The accuracy is again defined as the number of correct matches from the total samples. The reason for a higher accuracy compared with the balanced dataset is due to a lower fraction of early spirals in the imbalanced dataset ($\sim8\%$) than the balanced dataset ($\sim25\%$). When we exclude the early spirals from the imbalanced dataset, the accuracy barely changes, and it is consistent with the accuracy obtained when using the balanced dataset (accuracy: $\sim$0.87; Section~\ref{ch6_Sec:binary_clf_ba}). These results show the ability of our method to achieve reliable binary morphological classification for large surveys with unknown morphological mixes.


\subsubsection{Multiple classification clusters}
\label{ch6_Sec:ml_clf_im}

Following Section~\ref{ch6_Sec:clustering}, and using the imbalanced dataset, we obtain the same number of clusters, 27, as when we used the balanced dataset through our method of determining the number of clusters (Section~\ref{ch6_Sec:machine_clf}). The clustering results for both datasets are very close to each other, with only very minor differences. For example, 7 clusters are separated under the less featured group using the balanced dataset while 8 clusters are obtained using the imbalanced dataset. Conversely, we obtain 20 clusters for the featured group using the balanced dataset, and 19 using the imbalanced dataset.

\begin{figure*}
\begin{center}
\graphicspath{{figures/}}
	\includegraphics[width=2.1\columnwidth]{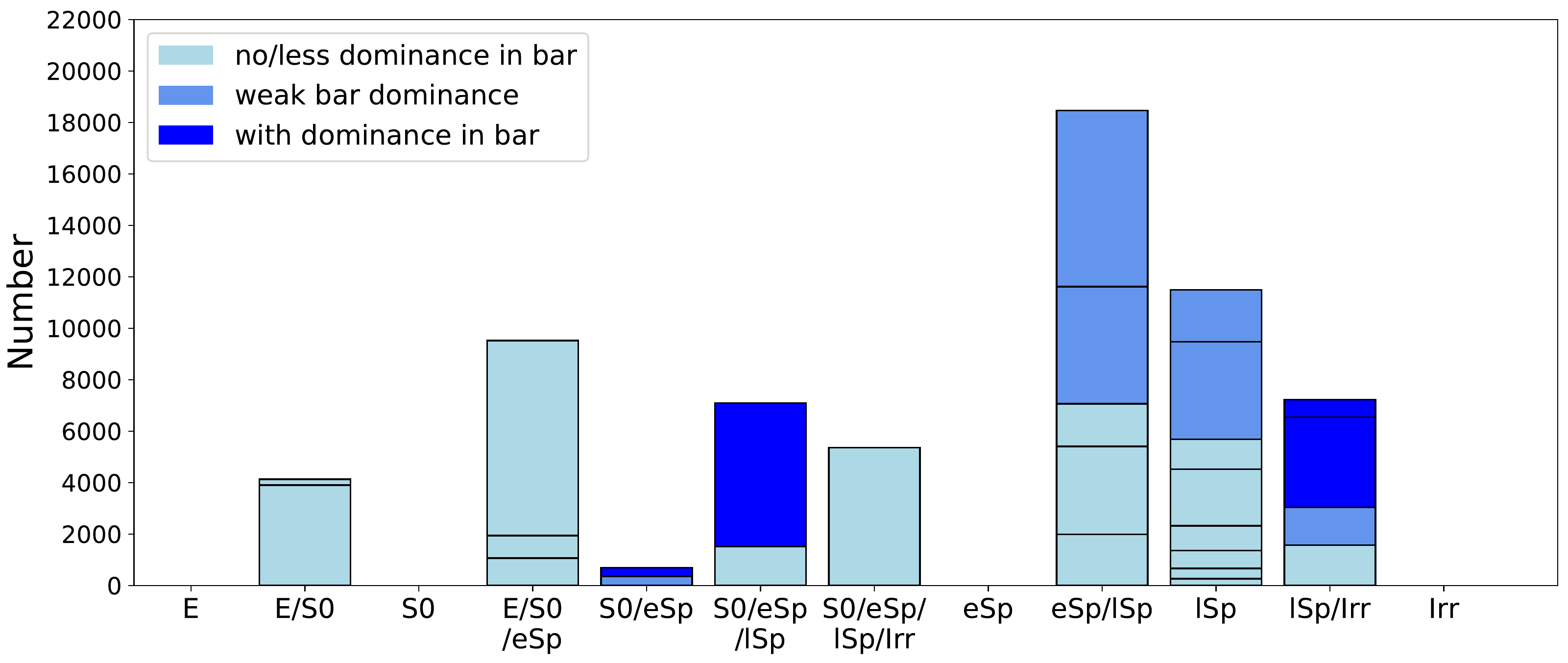}
   	\caption{The accumulated distribution of the classification clusters compared with the Hubble sequence for the imbalanced dataset. The $x$-axis shows one or a mix of visual morphology types. Each coloured bar represents one cluster. Different colours are different dominance of barred galaxies in the cluster, such that from deep to light blue represent more barred galaxies to no/fewer barred galaxies in the cluster.}
    \label{ch6_fig:ml_versus_HS_im}
\end{center}
\end{figure*}

In Fig.~\ref{ch6_fig:ml_versus_HS_im}, we associate the classification clusters for the imbalanced, realistic, data set with Hubble types based on the dominance of each type. We find no clean clusters for ellipticals (E), lenticulars (S0), early spirals (eSp), irregulars (Irr) when using the imbalanced dataset. The lack of clusters for $E$ and $eSp$ is due to the same reasons for the balanced dataset discussed in Section~\ref{ch6_Sec:machine_clf}: these two visual morphologies are intrinsically difficult to separated from other morphology types. Additionally, in Section~\ref{ch6_Sec:machine_clf}, we conclude that to get a clean $S0$ cluster, galaxies have to show a moderate disk structure (Fig.~\ref{ch6_fig:comp_sersic_E_S0_esp}). However, there is not a sufficient number of lenticulars with the relevant features due to the low fraction of this type in the imbalanced dataset (Fig.~\ref{ch6_fig:data_type_disb}). It is impossible for the machine to classify a galaxy that does not exist in some abundence within the dataset; therefore, we miss the pure $S0$ cluster when using the imbalanced dataset. On the other hand, irregular galaxies do not have a specific structure; therefore, it is easy to be confused them with some late spirals with less structured appearances by our machine, based on only galaxy structure and without the prior knowledge of `late sprials' or `irregulars'. They also suffer from the similar cause of the missing $S0$ cluster: the insufficient number of irregular galaxies in our imbalanced set decreases the possibility of the distinctive irregulars to be picked out by our machine.

Similar to the results of the balanced dataset, the separation between clusters might `improve' in terms of being closer to a more physical classification when we consider colour information in our machine. Therefore, this will be an important part in future work.

\section{Conclusions}
\label{ch6_sec:conclusion}

In this paper, we present an unsupervised machine learning technique by applying a combination of a feature extractor - a vector-quantised variational autoencoder (VQ-VAE) and a hierarchical clustering algorithm (HC). This method involves a vector quantisation process which provides a rate of classification with a feature extractor in the learning phase at least 30 times faster than a typical convolutional antoencoder used in \citet{Cheng2020b} on the same device.

To sensibly explore galaxy morphologies and investigate the suggestive number of galaxy morphological classes, we propose some novel modifications to the machine learning algorithms used in this work (Section~\ref{ch6_sec:methods}). First, we include a preliminary clustering result in the VQ-VAE architecture during the feature learning process. This helps to extract features that can not only reproduce the input images but also be well separated into two preliminary clusters in feature space. Second, different distance thresholds are used within each branch in the merger tree in the HC process rather than a single distance threshold for a whole tree. This flexibility prevents the creation of unnecessary clusters separating galaxies with few features, while allowing more clusters for galaxies that show larger variation. Another innovation is to use galaxy orientation (a potential problem when classifying galaxies) to our advantage, helping to decide the number of clusters (Section~\ref{ch6_Sec:clustering}).

Using the monochromatic images from the Sloan Digital Sky Survey (SDSS), we first explore galaxy classifications using a dataset with a balanced number of galaxies in each morphological class (Section~\ref{ch6_sec:datasets}).  This is done to reduce potential biases associated with number imbalances. Using this method we obtain 27 clusters within this balanced dataset. We find that our method separates the classification clusters based on galaxy shape and structure (e.g., S\'ersic index, asymmetry, Gini coefficient, M20). We then associate our classification clusters with the Hubble sequence based on the dominance of each type in a given cluster (Section~\ref{ch6_Sec:machine_clf}). Clusters with barred, weak-barred, and non-barred galaxies are well distinguished by our machine. However, when using the balanced dataset, no clean clusters are found for ellipticals or early spirals (Fig.~\ref{ch6_fig:ml_versus_HS_ba}). Additionally, most clusters are associated with a mixture of Hubble types. We thus conclude that there is a fundamental difficulty in separating accurately galaxies with transitional features such as lenticular galaxies and early spirals with a machine. This applies both to visual and machine classifications.

In addition, we find that each machine classification cluster has characteristic galaxy properties (e.g., colours, masses, luminosities, sizes) that transition smoothly along the Hubble sequence. In future work, we will further investigate if this unsupervised method could provide a more physically meaningful classification system than a purely visual classification one.

Overall, the machine classification clusters provide a reasonable and detailed scheme for galaxy morphological classification based on a combination of multiple structural parameters, avoiding human errors and biases. The dominate features in our classification clusters can be used as the foundation of an objective alternative to the Hubble sequence. Since our system separates well galaxies with different shape, structure, and physical properties, it may prove useful in generic galaxy formation and evolution studies. The system may be improved by including multi-colour imaging and velocity maps. Galaxies at higher redshifts have structures which look significantly different from those of nearby galaxies, such as those we examine in this study. Thus, it would be interesting to apply our technique to higher redshift galaxies to see whether the VQ-VAE method would classify galaxies in the earlier universe into our clusters or if it would suggest new ones.

To test the performance of our method with realistic morphological distributions, we also apply it to an imbalanced dataset which follows the morphological distribution of nearby galaxies. The results are very similar to the ones obtained with the balanced dataset, showing that our system is able to deal with large galaxy samples with more realistic morphological mixes.  It also shows that our set up is not sensitive to different distributions of input galaxy morphologies, but can handle a range of distributions of various galaxy input `types'.

As mentioned earlier, in the future we plan to carry out a more detailed comparison between a machine-defined classification and a visual classification to investigate the pros and cons between the two ways. In addition, we intend to apply the techniques developed here to multi-colour images with better resolution such as the data from the Dark Energy Survey and the Euclid Space Telescope. Velocity maps from integral-field spectroscopic surveys could also be included. The resulting classification system(s) should prove very useful to better understand galaxy properties, their formation and evolution. We also expect that the future development of this work will result in a fundamental change in how we approach galaxy morphological classification - both visually and when using machine learning.

\section*{Acknowledgements}
This work was initiated for the Kavli Summer Program in Astrophysics held at the University of California Santa Cruz (UCSC) in 2019. T.-Y. Cheng thanks the organisers and UCSC for the great opportunity and generous financial support during the summer program. Thanks to Francois Lanusse for a great discussion when this project was initiated, and Asa F. L. Bluck for advising appropriate catalogues for validation. Additionally, T.-Y. Cheng acknowledges the support of the Vice-Chancellor's Scholarship from the University of Nottingham. The authors based in UK acknowledge the support by the UK Science and Technology Facilities Council (STFC). NR's work at Argonne National Laboratory was supported by the U.S. Department of Energy, Office of High Energy Physics. Argonne, a U.S. Department of Energy Office of Science Laboratory, is operated by UChicago Argonne LLC under contract no. DE-AC02-06CH11357. BER acknowledges support from NASA contract NNG16PJ25C and grant 80NSSC18K0563, and NSF grant AST 1828315.

\section*{Data Availability}
The simplified version of code used in this work is published on GitHub - https://github.com/tycheng-sunny/Unsupervised-ML-on-Galaxy-Morphology-using-VQ-VAE. The complete codes and data underlying this article can be shared on reasonable request to the corresponding author. The raw imaging data can be obtained from the SDSS DR7 archive - http://cas.sdss.org/dr7/en/tools/crossid/upload.asp.

\bibliographystyle{mnras}
\bibliography{ms}


\label{lastpage}
\end{document}